\documentclass[prd,amsmath]{revtex4}
\usepackage{graphicx}
\usepackage{hyperref}
\usepackage{bm}
\usepackage{natbib}
\usepackage{ulem}

\usepackage{float}
\usepackage{subfig}
\usepackage{amssymb}

\renewcommand{\vec}[1]{\bm{#1}}

\def\o3z{$O(3)_z$}

\begin{document}
\title{Numerical Exploration of Soliton Creation}

\author{Henry Lamm, Tanmay Vachaspati}
\affiliation{Physics Department, Arizona State University, Tempe, AZ 85287}

\begin{abstract}
We explore the classical production of solitons in the easy axis O(3) model in 1+1 dimensions, 
for a wide range of initial conditions that correspond to the scattering of small breathers. We 
characterize the fractal nature of the region in parameter space that leads to soliton production and find certain trends in the data. We identify a tension in the initial conditions required for soliton 
production --  low velocity incoming breathers are more likely to produce solitons, while 
high velocity incoming breathers provide momentum to the final solitons and enable them to
separate. We find new ``counter-spinning'' initial conditions that can alleviate some of this
tension.
\end{abstract}

\maketitle

\section{Introduction}
\label{sec:introduction}

Perturbative field theory is an expansion around a single vacuum state but many field theories admit
multiple degenerate vacua. Excitations in such field theories include solitons that interpolate between different vacua in addition to the particle excitations around a single vacuum. While solitons and particles are distinct excitations, it is generally possible to transition from one to the other. For example, a soliton and anti-soliton are able to annihilate and produce particles. Here we shall be concerned with the reverse process in which we start with particles and create a soliton-antisoliton pair.

The transition from particles to solitons (or vice versa) has another layer of complexity because
it is also a transition from a quantum system (particles) to a classical system (solitons). A rigorous
formalism to treat the transition is not known. While some attempts have been made at semiclassical and quantum calculations \cite{Mottola:1988ff,Ringwald:1989ee,Mattis:1991bj,Rebbi:1996zx,Kuznetsov:1997az,Bezrukov:2003er,Levkov:2004tf,Demidov:2011dk}, the most
straightforward approach at the current time is to 
treat the entire process classically. Particles in the initial state are replaced by classical
field configurations that are non-dissipative, like ``breathers'' of the sine-Gordon 
model \cite{Vachaspati:2011ad}, or dissipative but long-lived,
like ``oscillons" \cite{Dutta:2008jt,Romanczukiewicz:2010eg,Demidov:2011eu} in other models.

Previous work on the classical production of solitons has mostly been carried out in the
$\lambda\phi^4$ model, which has the virtue that it has the minimal structure necessary 
for studying the process \cite{Dutta:2008jt,Romanczukiewicz:2010eg,Anninos:1991un,Campbell:1983xu}. However, the simplicity of the model may also
be a drawback, as additional degrees of freedom  \cite{Vachaspati:2011ad} or a more complex potential \cite{Demidov:2011eu,Demidov:2011dk} may facilitate the production of
solitons. Thus we study soliton production in the easy axis
O(3) model (described in detail in Sec.~\ref{sec:model}).

The easy axis O(3) model (or ``\o3z model'') in 1+1 dimensions has a 
number of features that make it suitable for studying soliton production. As the model has two 
degenerate vacua, it contains kink solutions. Certain subspaces of the model are equivalent 
to the classical sine-Gordon model. Thus the model also has breather solutions that do not
decay and can be used to mimic incoming particle states. The \o3z model has an additional
``twist'' degree of freedom that gives it more complexity than the $\lambda\phi^4$ model
and brings it a bit closer to models with `t Hooft-Polyakov magnetic monopoles, as monopoles
also carry a phase degree of freedom. 

This paper is organized as follows.  In Sec.~\ref{sec:model} we describe the \o3z model and
in Sec.~\ref{sec:in} we describe the range of initial conditions that we use in our scattering
simulations. Our numerical results are discussed in Sec.~\ref{sec:numres} and
we conclude in Sec.~\ref{sec:con}.

\section{Easy axis $O(3)$ Model}
\label{sec:model}

The \o3z model  in 1+1 dimensions is given by the action
\begin{equation}
 S = \int d^2x \left [ \frac{1}{2}(\partial_\mu {\vec n})^2 - \frac{1}{2}(1-n^2_3)
                      - \lambda ({\vec n}^2 -1)
               \right ]
\label{model}
\end{equation}
where ${\vec n}(t,x)$ is a vector field with Cartesian components $(n_1,n_2,n_3)$
and $\lambda$ is a Lagrange multiplier that forces ${\vec n}$ to have unit
magnitude: ${\vec n}^2 = 1$. The potential term reduces the $O(3)$ symmetry to
$O(2)\times Z_2$, corresponding to symmetry under  rotations in the $n_1$-$n_2$
plane and to reflections of $n_3$. There are two degenerate vacua: ${\vec n}=(0,0,\pm 1)$.

After eliminating the constraint condition, the equation of motion is
\begin{equation}
\Box {\vec n} + (\partial_\mu {\vec n})^2 {\vec n}
      - n_3 ( {\hat e}_3 - n_3 {\vec n} ) =0 \ ,
{\vec n}^2 = 1
\label{nevolve}
\end{equation}
where ${\hat e}_3 \equiv (0,0,1)$.

Alternately, the constraint can be solved explicitly in terms of angular variables,
${\vec n}= (\sin\theta\cos\phi,\sin\theta\sin\phi,\cos\theta)$,
to give the action
\begin{equation}
 S = \int d^2x \left [ \frac{1}{2}(\partial_\mu \theta)^2+
         \frac{1}{2}\sin^2\theta (\partial_\mu \phi)^2-\frac{1}{2}\sin^2\theta
              \right ]
\end{equation}
which leads to the equations of motion
\begin{eqnarray}
\label{eom}
\Box\theta+\sin\theta \cos\theta (1-(\partial_\mu\phi)^2) &=&0\\
\partial_\mu(\sin^2\theta ~ \partial^{\mu}\phi) &=&0.
\end{eqnarray}
The latter equation is of the form $\partial_\mu j^\mu =0$ and so $j^0=\sin^2\theta {\dot \phi}$
is the charge density of a conserved current in the model.

\begin{figure}
\begin{center}
\includegraphics[width=4 in]{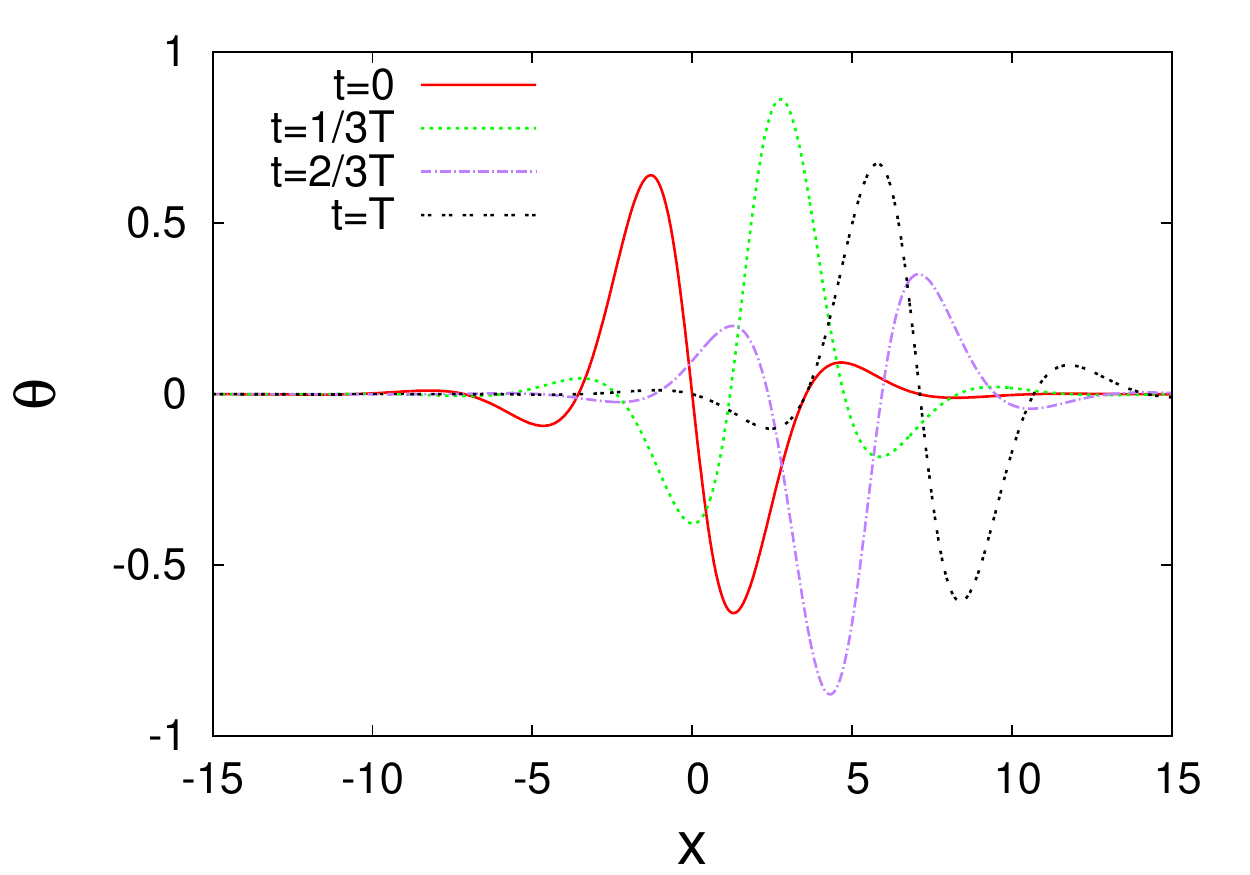}
\caption{Field profile ($\theta$) of a boosted breather at different times over an oscillation
period. The particular parameters used are $v=0.9$, $\omega=0.90$ which corresponds to $m_b=1.74$.
\label{fig:kinkbreather}
}
\end{center}
\end{figure}

Let $\alpha\equiv2\theta$ and consider the case, $\phi={\rm constant}$. Then the
equation of motion reduces to
\begin{equation}
 \Box\alpha+\sin\alpha =0
\end{equation}
which is identical to the equation of motion for the sine-Gordon model. (This can
also be seen at the level of the action.) Hence the \o3z model inherits
all the solutions of the sine-Gordon model. In particular
\begin{equation}
 \theta_k(x)=2\tan^{-1}(e^x)\ , \ \ \phi={\rm constant}
\end{equation}
is a kink solution in which ${\vec n}$ is at the North pole in field
space at $x=-\infty$ and at the South pole at $x=+\infty$.
With our normalization, the energy of the kink is $E_k=2$.
The model also inherits the (boosted) sine-Gordon breather solutions
\begin{equation}
 \theta_b (x,t) = 2\tan^{-1}\left[\frac{\eta \sin(\omega T)}{\cosh(\eta \omega X)}\right],
 \ \  \phi = {\rm constant}
 \label{thetab}
\end{equation}
where 
\begin{equation}
 T = \gamma(t-v(x-x_0))
\end{equation}
\begin{equation}
 X = \gamma((x-x_0)-vt)\ .
\end{equation}
The parameter $x_0$ is the initial position of the breather, $v$ its velocity,
and $\gamma=1/\sqrt{1-v^2}$ its Lorentz factor. The parameter $\omega$ is the oscillation
frequency of the breather and takes values in $[0,1]$, while $\eta$ is
defined by
\begin{equation}
 \eta = \sqrt{1-\omega^2}/\omega
\end{equation}
The typical waveform of a boosted breather can be seen in Fig.~\ref{fig:kinkbreather}.  

The energy in a field configuration is given by
\begin{widetext}
\begin{equation}
E = \int dx~ H(t,x) = \int dx \left [
   \frac{{\dot\theta}^2}{2} +
   \frac{{\theta '}^2}{2} +
   \sin^2\theta\frac{{\dot\phi}^2}{2} +
   \sin^2\theta\frac{{\phi'}^2}{2} +
   \frac{1}{2}\sin^2\theta \right ]
\label{hamiltonian}
\end{equation}
\end{widetext}

For a boosted breather, the energy evaluates to
\begin{equation}
 E_b = 4\gamma\sqrt{1-\omega^2}=4\gamma\eta\omega
\label{eq:energyb}
\end{equation}
and the rest mass of the breather is
\begin{equation}
 m_b=\frac{E_b}{\gamma}=4\sqrt{1-\omega^2}=4\eta\omega
\end{equation}
Note that the energy of the breather is smaller for larger $\omega$ and 
vanishes for $\omega =1$.
If we choose initial conditions on a great circle ($\phi = \phi_0, \phi_0+\pi$), the 
dynamics will also be restricted to the great circle. Then the evolution is exactly as in
the sine-Gordon model. 
As the sine-Gordon model is completely integrable \cite{Zamolodchikov:1978xm},
both the number of kinks and the number of anti-kinks are conserved and we can 
not create kinks (or antikinks) if there were none in the initial conditions. 
Hence it is crucial to 
choose initial conditions in which $\phi$ is not a constant {\it i.e.}
that the initial conditions be ``twisted'' in the $\phi$ direction.

A general feature in 1+1 classical field theory of interest to us will be that the 
interaction force between two static solitons separated by a distance $L$
is proportional to
\begin{equation}
\label{eq:forceg}
 F(L)\propto e^{-L/a}
\end{equation}
where $a$ is a length scale, usually on the order of the width of a soliton
\cite{Vachaspati:2006zz}.

\section{Choice of initial conditions}
\label{sec:in}

Ideally we would like to start with initial conditions that describe incoming
particles in some energy range, but then the initial condition would have
to be described in terms of quantum field theory. On the other hand, the
final state of interest has kinks, and these are classical objects.
So the creation of kinks from particles also involves a transition from an
initial quantum system to a final system that contains classical elements, 
and a formalism to describe such a transition is not known. The simplification
we will adopt is to consider initial conditions that only contain breathers as,
at least in the sine-Gordon model, it has been shown that quantized breathers
correspond to particles of the theory in the low mass limit \cite{Dashen:1975hd}.

Our choice of initial conditions consists of a train of $N$ left-moving
breathers and $N$ right-moving breathers. The parameters $m_b$ and
$v$ characterize an individual breather in the train. In addition, we can
vary the spacing of the breathers within a train, the number of breathers
in each train, and the relative twists of the breathers. This last parameter
is called $\xi$ and is defined by
\begin{equation}
\xi \equiv \frac{\phi_L-\phi_R}{2\pi}
\end{equation}
where $\phi_L$, $\phi_R$ are the initial constant values of $\phi$ for 
the left- and right-moving breathers.
Explicitly, the initial condition is
\begin{eqnarray}
\theta(t=0,x) &=& \sum_{j=1}^{N}
    \left [ \theta_b(t=0,x;-x_j,+v) + \theta_b(t=0,x;+x_j,-v) \right ]
\label{Nbic}\\
\phi(t=0,x) &=& \pi \xi ~  {\rm tanh}(x/w) \\
{\dot \theta}(t=0,x) &=& \sum_{j=1}^N
    \left [{\dot \theta}_b(t=0,x;-x_j,v) +
                           {\dot \theta}_b(t=0,x;+x_j,-v) \right ] \\
{\dot \phi}(t=0,x) &=& 0
\label{eq:Ndotphi}
\end{eqnarray}
where $\theta_b$ is the breather profile defined in Eq.~(\ref{thetab}) and the initial position
 of the $j^{\rm th}$ breather given by
\begin{equation}
x_j = x_0 + (j-1) a .
\end{equation}
Here $x_0$ is the initial position of the innermost breather in the train
and $a$ is the spacing between the breathers in a train. 
In our numerical runs, the parameter $a$ is chosen to be twice the width of a breather, 
$a = 2w = 4/(\gamma\eta\omega )$, and $x_0$ is chosen to be $4 E_b w$ 
which is much larger than the width of the breathers investigated.
In the bulk of our analysis, we start with $\dot\phi=0$, but in
Sec.~\ref{sec:initialvel} we will also describe some results with initial conditions
${\dot \phi}(t=0,x)\ne 0$.

The initial conditions listed above are used to construct the vector 
${\vec n}(t=0,x)$, which is then numerically evolved using Eq.~(\ref{nevolve}).
The numerical evolution is done using the explicit second order Crank-Nicholson 
method with two iterations \cite{Teukolsky:1999rm}.

We wish to explore a large number of initial conditions and to record
only those initial conditions that lead to kink formation. Hence we need 
to specify criteria to decide if kinks were or were not produced in any given 
run. To do this, for each time step, we checked for a transition from $\cos\theta=0$ to $\cos\theta=-0.99$.  If this transition exists, the point where $\cos\theta=0$ is considered to be the location of the kink at that time.  By recording kink locations as a function of time, we were able to 
reconstruct the kink's path and therefore its velocity.
To explore parameter space, we hold $x_0$ and $a$ fixed and scan over a range of the 
4 parameters: $\omega$ (breather frequency), $v$ (incoming velocity), $\xi$ (twist),
and $N$ (number of breathers in a train).  To search this phase space for successful kink
production, we used two different methods.  For coarse grained searches, we used the MULTINEST software to find large clusters of conditions that lead to success \cite{Feroz:2008xx,Feroz:2007kg}.  For fine grained searches at low mass, we scanned the 
initial conditions uniformly in steps of $\Delta v=0.002$, $\Delta \xi=0.001$ and $\Delta \omega = 0.01$.  The ranges we explore are
shown in Table~\ref{tab:param}. We also show the corresponding range 
of the mass ($m_b$) and the Lorentz factor
$\gamma$. ($m_b$ and $\gamma$ are derived from the ranges of $\omega$ 
and $v$.)

\begin{table}[ht]
\begin{tabular}{l|c}
\hline\hline
Parameter & Range\\
\hline
$\omega$ & 0.500-0.990 \\
$v$ & 0.100-0.990\\
$\xi$ & 0.000-0.500\\
$N$ & 1-20\\
\hline
$m_b$ & 0.565-3.500 \\
$\gamma$ & 1.005-7.088\\
\hline
\end{tabular}
\caption{Range of initial conditions that we explore. For reference,
the rest energy of a kink is 2.}
\label{tab:param}
\end{table}

We now describe our numerical results.

\section{Numerical results}
\label{sec:numres}

\begin{widetext}
\begin{figure}
\begin{center}
\raisebox{.05\linewidth}{\subfloat{\includegraphics[width=0.17\linewidth]{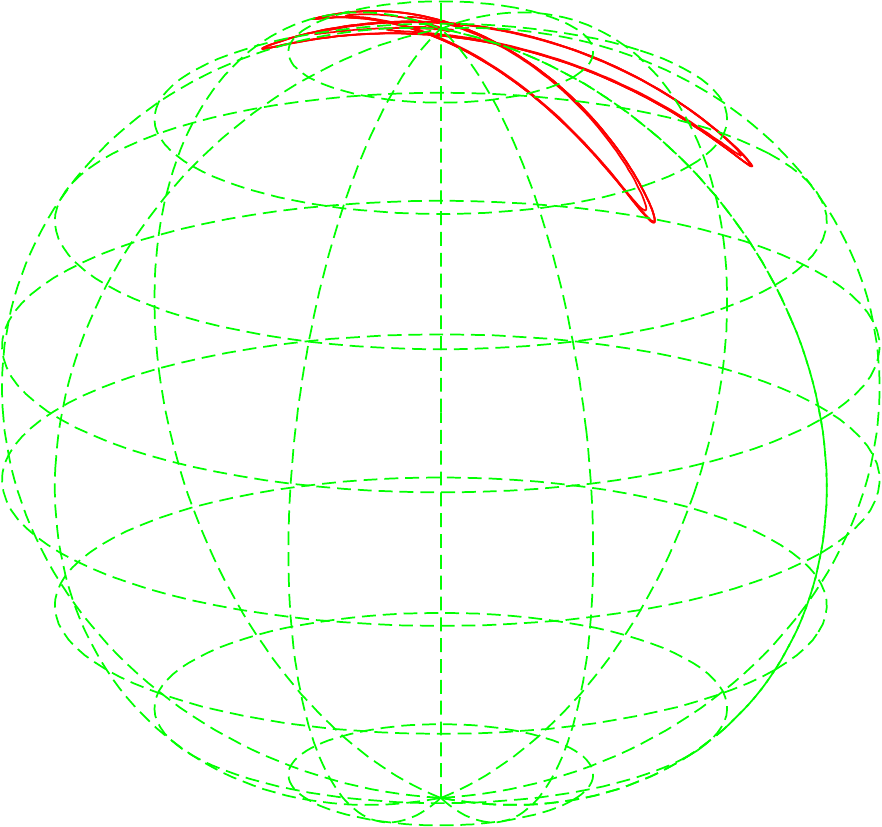}}}\,\,
 \subfloat{\includegraphics[width=0.6\linewidth]{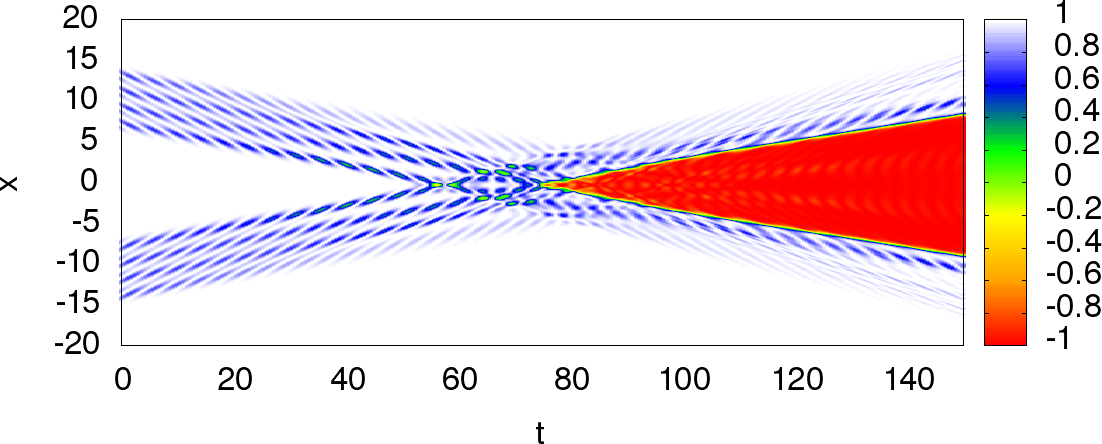}}\,\,
 \raisebox{.05\linewidth}{\subfloat{\includegraphics[width=0.17\linewidth]{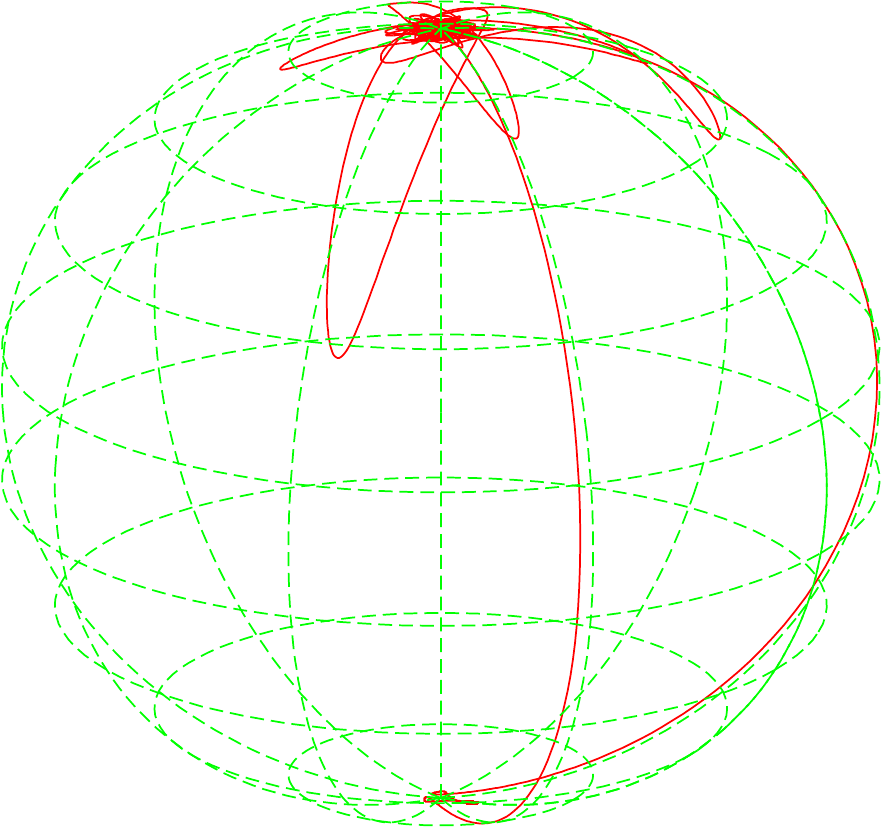}}}
\end{center}
\caption{The center panel indicates in color the $n_3$ field as a function of 
time (horizontal axis) and space measured in units of breather width(vertical axis) for the collision.  Left and right panels are the vector field ${\vec n}$ configurations at $t=0$ and $t=140$ respectively. The parameters for this scattering are $N=4$, $m_b=1.57$, $v_b=0.55$, and $\xi=0.1$. The final state contains 
a kink-antikink pair located at $x\approx\pm6.4w$, where $w$ is the
width of the incoming breather.}
\label{fig:focus}
\end{figure}
\end{widetext}

In Fig.~\ref{fig:focus} we show a sample event where a kink-antikink pair
is produced.  In the center we have the time evolution of the $n_3$ field.  One should note that while the breathers are initially moving with the same velocity, interactions within a train change the separation and relative velocity between breathers. 
The importance of this effect will be discussed in Sec.~\ref{sec:num}. 
The vector field ${\vec n}(t,x)$ is 
shown to the left and right of Fig.~\ref{fig:focus} at the initial and final times respectively.

\begin{figure}[ht]
\begin{center}
\subfloat{\includegraphics[width=0.5\linewidth]{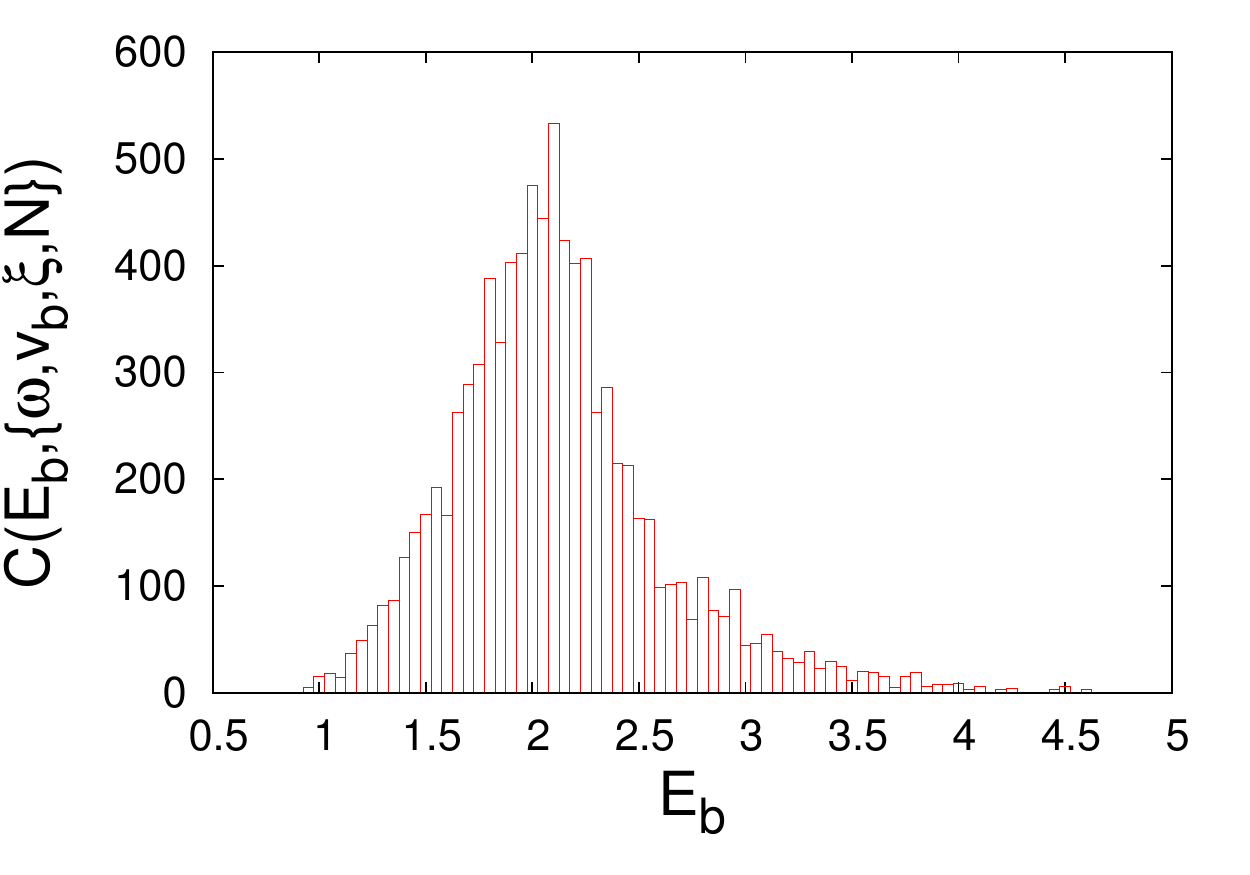}}
\subfloat{\includegraphics[width=0.5\linewidth]{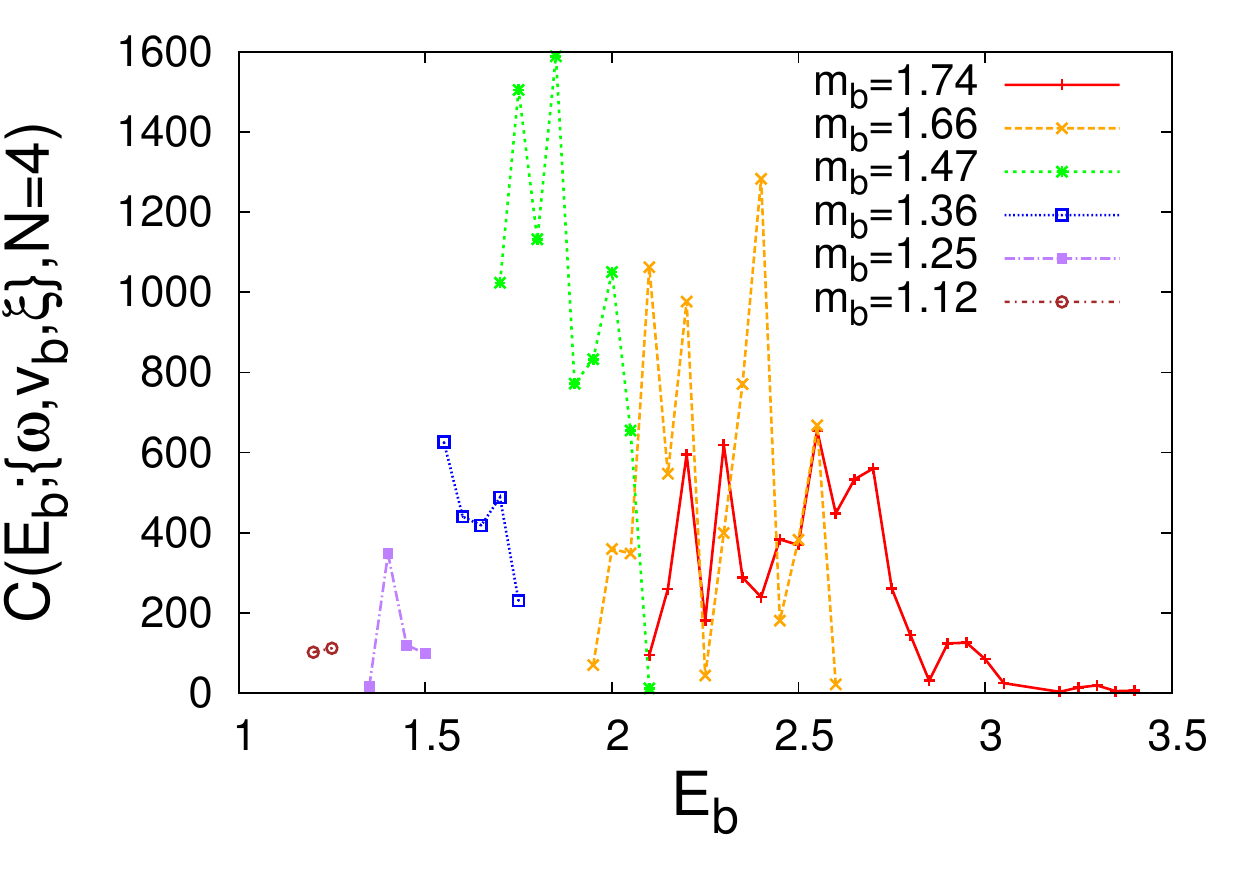}}
\caption{\label{fig:NvE}
Number of produced kinks vs. the energy of a single
breather, summed over $m_b,v_b,\xi,N$ for entire dataset.  A strong peak
is seen around the mass of a kink, $E_k=2$.  In the right-hand
figure, we plot the number of produced kinks vs. energy of a single 
breather, split into groups by breather mass, summed
over $v_b,~\xi$ but with $N=4$.}
\end{center} 
\end{figure}

We start by showing the cumulative result from all runs in the left panel of Fig.~\ref{fig:NvE}.  Plotted is the number of successful kink production events versus the energy per incoming breather, after summing over all other parameters in the 
ranges shown in Table~\ref{tab:param}. These results shows a peak in
production when the energy of an individual breather in the train is exactly 
the kink energy. This suggests that kink production is dominated by the collision
of just two breathers scattering into a kink-antikink pair. However, we will show that multi-particle interactions within the train are critical to the success of these collisions and that having 
$E_b\approx 2$ is not a sufficient condition for kink production. 

In the right panel of Fig.~\ref{fig:NvE}
we further partition the data of the left panel of Fig.~\ref{fig:NvE}
by the mass of the incoming breather but only for $N=4$. 
In this plot we see that there is a distribution of energies for which kink production
occurs, but the distribution is more narrowly distributed at smaller masses.
In other words, if the incoming breather has small mass, it's energy must be
picked more precisely.
While the general distribution of energies seems to be smooth and peaked around $E_b=2$, for a given mass of a breather there is a structure of clear peaks where success occurs, and a slight 
change in the energy can dramatically changes the success rate. 

\begin{widetext}
\begin{figure}
\begin{center}
\includegraphics[width=\linewidth]{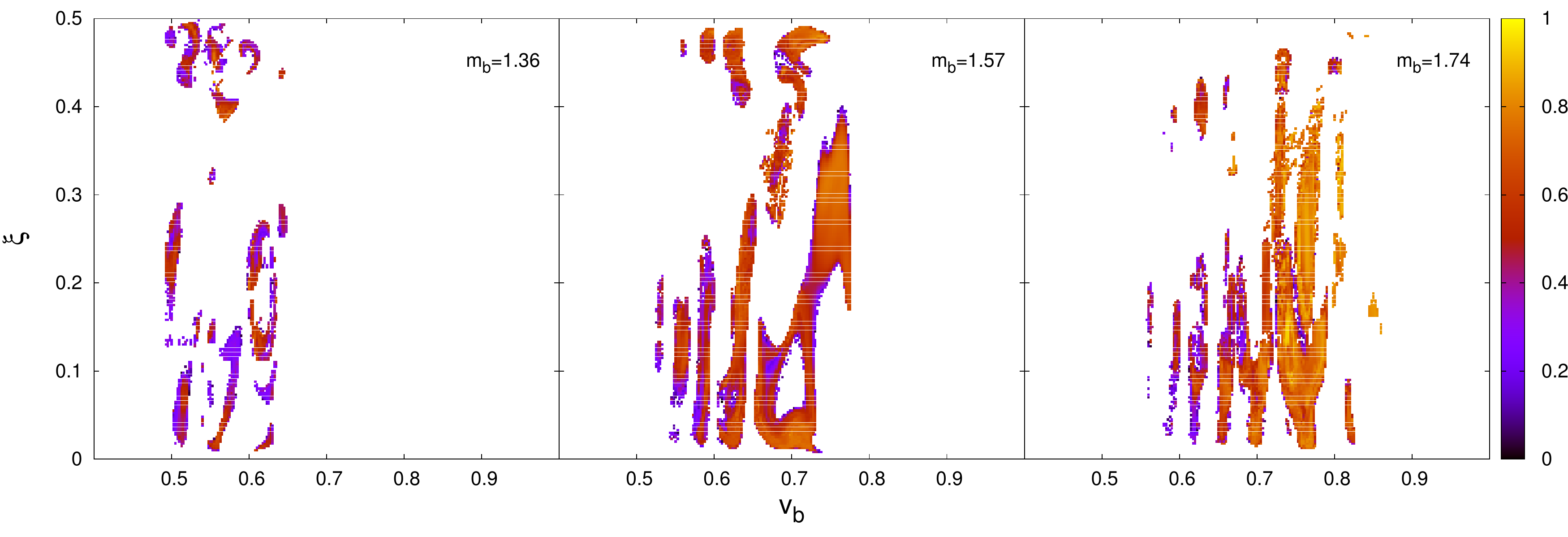}
\caption{\label{fig:kv} Velocity of outgoing kinks (denoted by color) as a function 
of incoming breather velocity and twist for three breather masses.  
Band structure can be seen in the velocity dependence of successful 
production.  The distinct drop in
production for $\xi= 0.25-0.40$ can be observed for all masses. Further, we observe that lower values of $m_b$ produce lower kink velocities.}
\end{center}
\end{figure}
\end{widetext}

Fig.~\ref{fig:NvE} provides a summary of all our runs but loses
information about the effects of varying parameters on kink
production. In order to untangle some of these effects, we 
investigated the case of $N=4$.  This condition was chosen as a balance 
between the increased number of successes that comes from having larger $N$, and 
the simpler dynamics afforded by
the few-body collisions of small $N$.  For $N=4$, we were able to find hundreds
of successful initial conditions. From these results, we computed the velocity
of the outgoing kinks as a function of the initial conditions.  For three
values of $m_b$, the outgoing kink velocities are plotted in 
Fig.~\ref{fig:kv}.  

A few distinct features can be seen in Fig.~\ref{fig:kv}. With
increasing breather mass, we see that the likelihood of producing
kinks increases, but the range of velocities that yield kinks in the 
final state shift to higher values.
Breather velocities where solitons are produced are found to form bands,
reminiscent of \cite{Campbell:1983xu} where the annihilation of
solitons into particle-like states was considered.  
The chaotic results -- note the hole at $v_b=0.7$,
$\xi=0.1$ in the middle panel -- also bear a
qualitative resemblance to the production found in
\cite{Romanczukiewicz:2010eg}.

\begin{figure}
\begin{center}
\includegraphics[width=4 in]{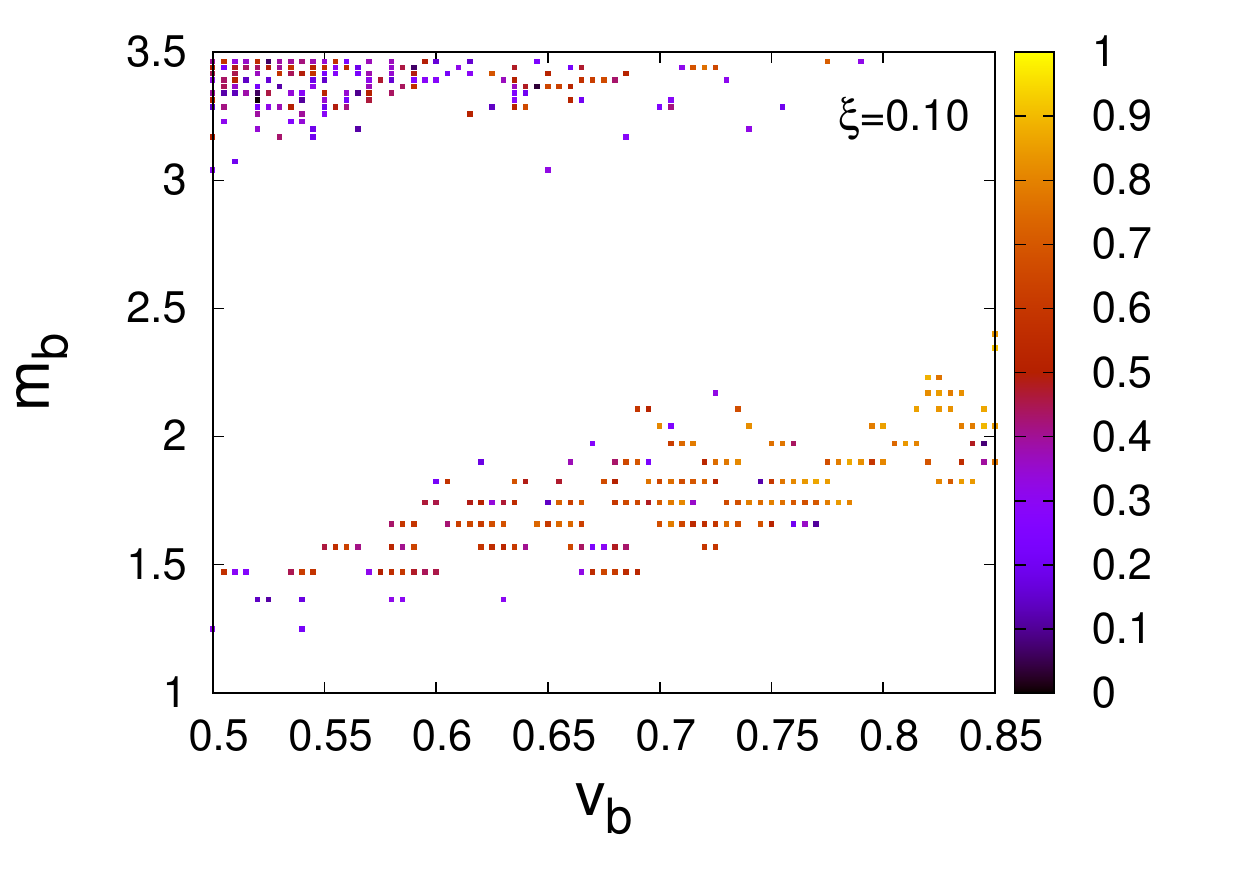}
\caption{\label{fig:kvvk} $v_k$ (denoted by color) is plotted vs $v_b$ and $m_b$ for $\xi=0.10$.
Successful kink production occurs in two bands that are well separated.}
\end{center}
\end{figure}

In these plots, the likelihood of kink production is suppressed in a 
region around $\xi\approx0.25-0.35$ except at higher velocities where kink production is relatively insensitive to changes in twist.
Another interesting feature is found by considering the dependence of
clustering in the ($m_b,v_b$) plane.  In Fig.~\ref{fig:kvvk}, we plot 
the outgoing kink velocity for $\xi=0.10$. 
Notice the counter-intuitive trend that for decreasing breather
mass, $m_b$, successful kink production requires a {\it decrease} in the
incoming velocity. This is the same dependence found in the 
$\lambda\phi^4$ model \cite{Dutta:2008jt} and suggests a
difficulty in the production of solitons from quantum particles as we will discuss below.

Also of interest in Fig.~\ref{fig:kvvk} is the appearance of two bands, one at
high $m_b$ and another at low $m_b$.  Inside of each band, we see that increasing the 
breather mass requires an increase in velocity for kink production to be successful. 
The optimal velocity for kink production is a function of breather mass, and
it increases within a band, but then has a large discontinuous jump as we cross from 
one band to another.

\begin{widetext}
\begin{figure}
\begin{center}
\subfloat{\includegraphics[width=0.33\linewidth]{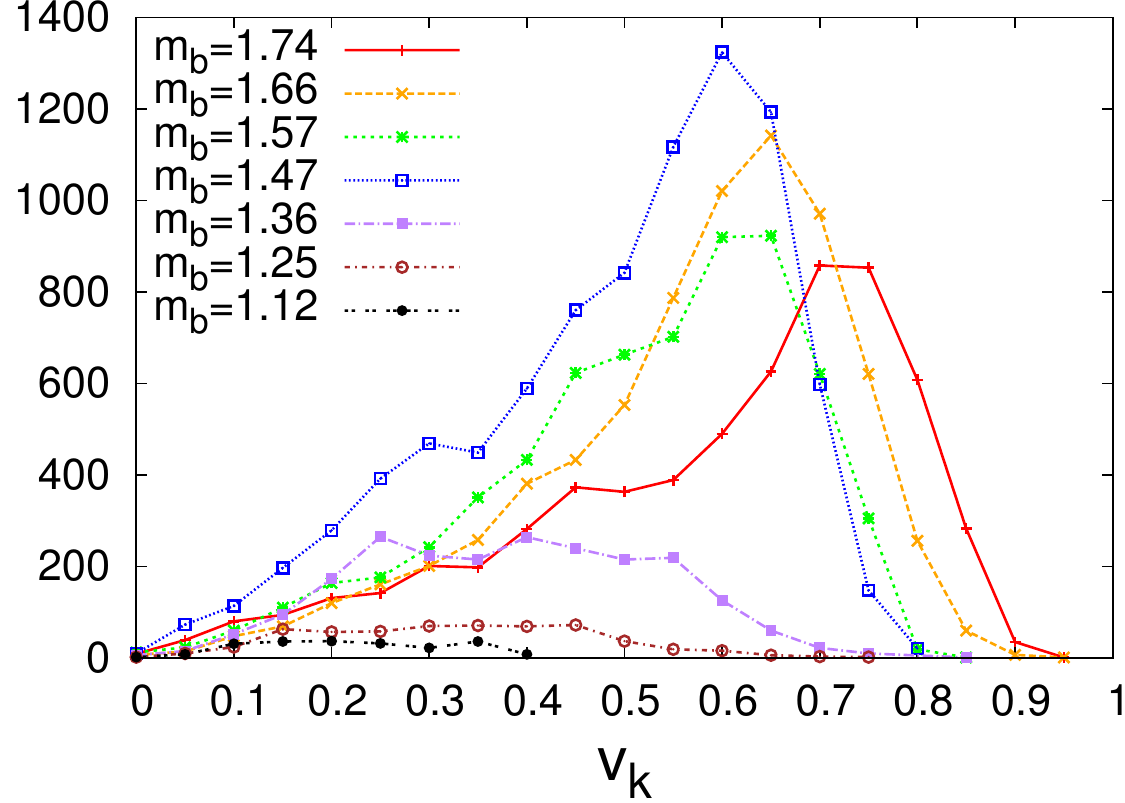}}
\subfloat{\includegraphics[width=0.33\linewidth]{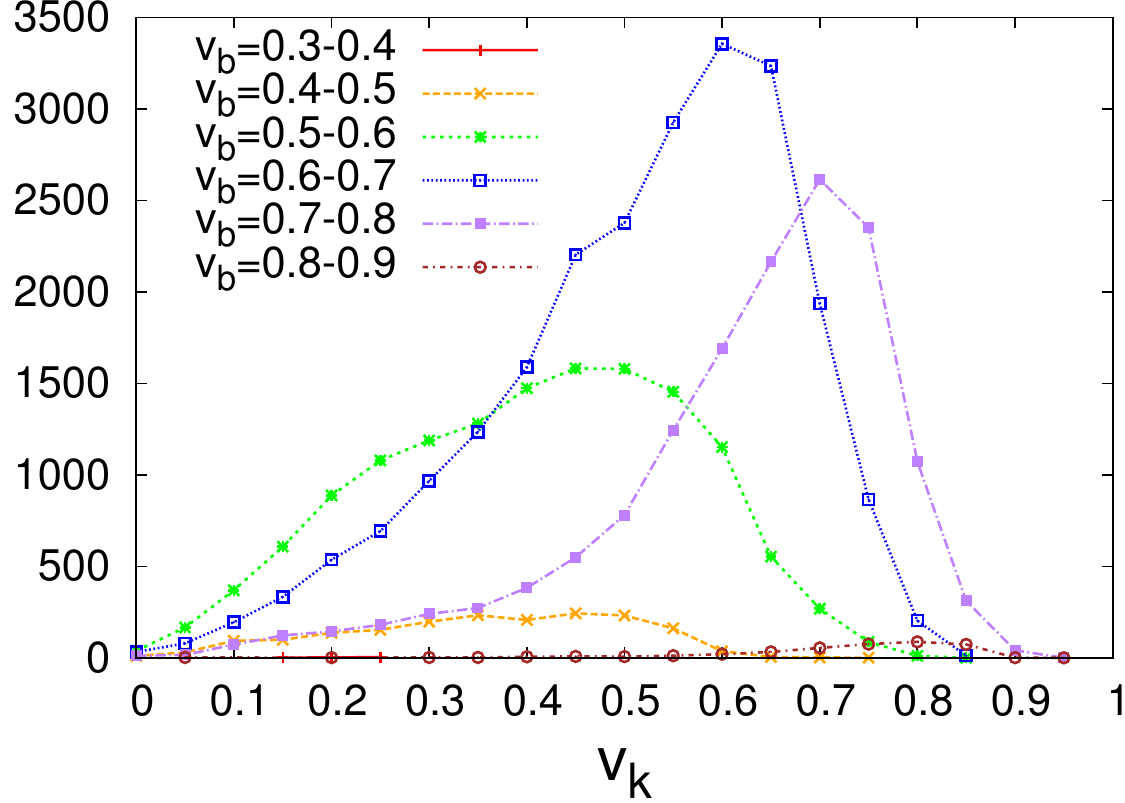}}
\subfloat{\includegraphics[width=0.33\linewidth]{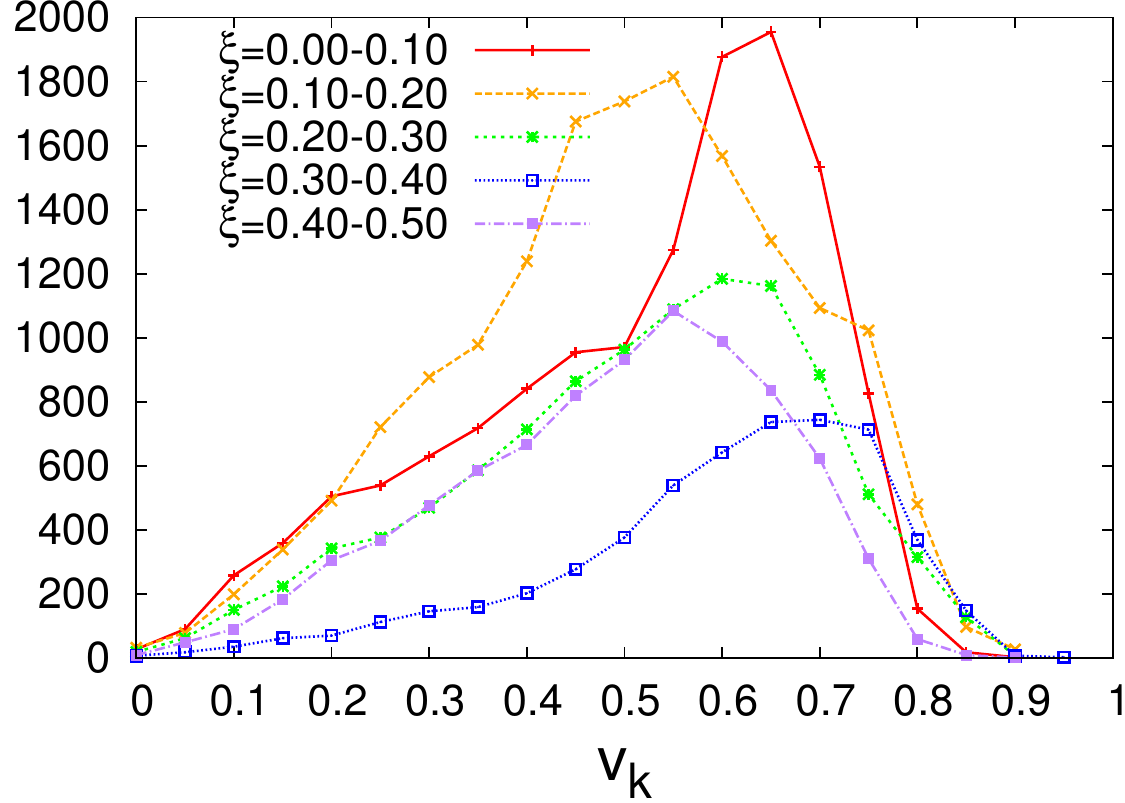}}
\caption{\label{fig:vkvsparams}
Cumulative number of kink production events produced with velocity $v_k$ from the set of initial conditions $m_b \in (1.12,1.74)$, $v_b \in (0.0,1.0)$, and $\xi \in (0.0,0.5)$. In the left panel, successful events for all $v_b$ and $\xi$ are accumulated; similarly
in the central panel, events for $m_b$ and $\xi$ are accumulated,
and in the right panel, events for $m_b$ and $v_b$ are accumulated.}
\end{center}
\end{figure}
\end{widetext}

Next we focus on the kink velocity, $v_k$, in the runs that successfully
produced a kink-antikink pair. Fig.~\ref{fig:vkvsparams} shows the dependence 
of $v_k$ on the parameters $m_b$, $v_b$ and $\xi$, where we fix $N=4$. For example,
the left-most panel shows that for $m_b=1.74$ and $N=4$, most of the successful
runs produced kinks with velocity $\approx 0.7$. As we decrease $m_b$, the velocity of the outgoing kinks decreases {\it e.g.} at $m_b=1.47$, the peak occurs at $v_k \approx 0.6$. At $m_b=1.12$, the success rate for kink productions very low and the outgoing kink velocity has decreased to $\approx 0.3$. Similarly, the center and right-hand panels of Fig.~\ref{fig:vkvsparams} show the dependence of $v_k$ on $v_b$ and $\xi$.

\begin{widetext}
\begin{figure}
\begin{center}
\subfloat[$\langle v_k\rangle_{v_b,\xi}$ vs.
$m_b$]{\includegraphics[width=0.33\linewidth]{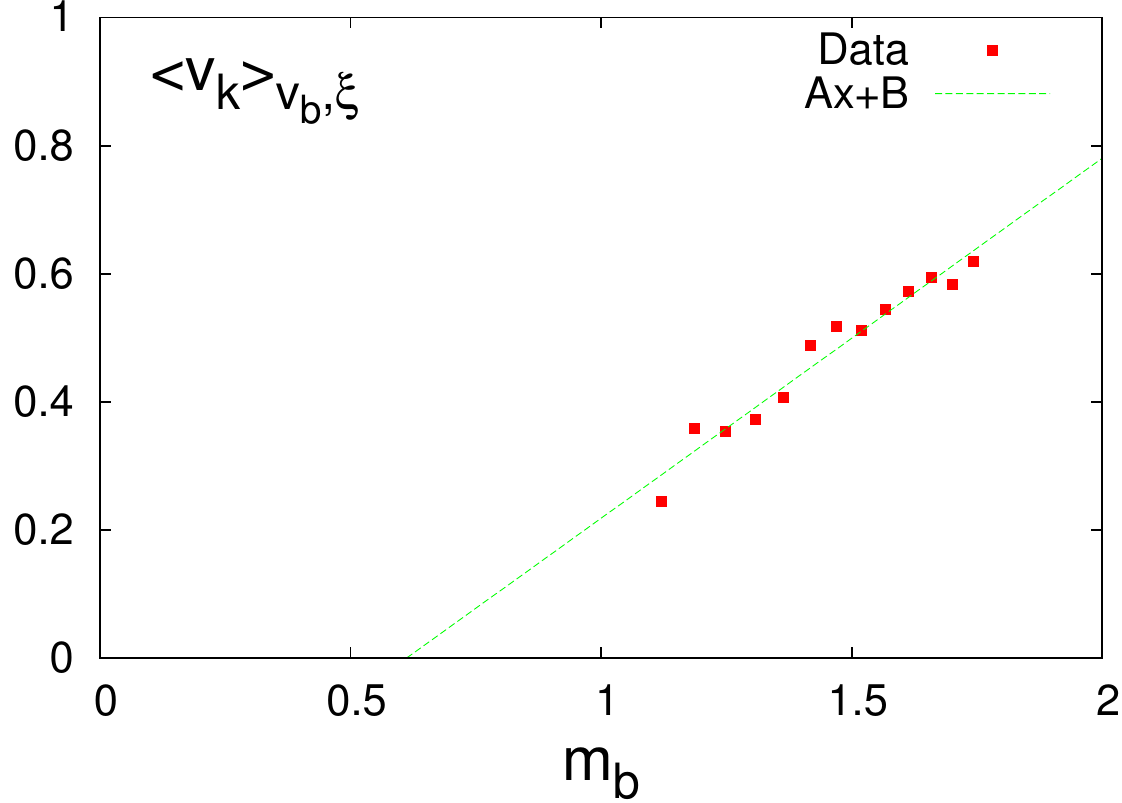}}
\subfloat[$\langle v_k\rangle_{m_b,\xi}$ vs.
$v_b$]{\includegraphics[width=0.33\linewidth]{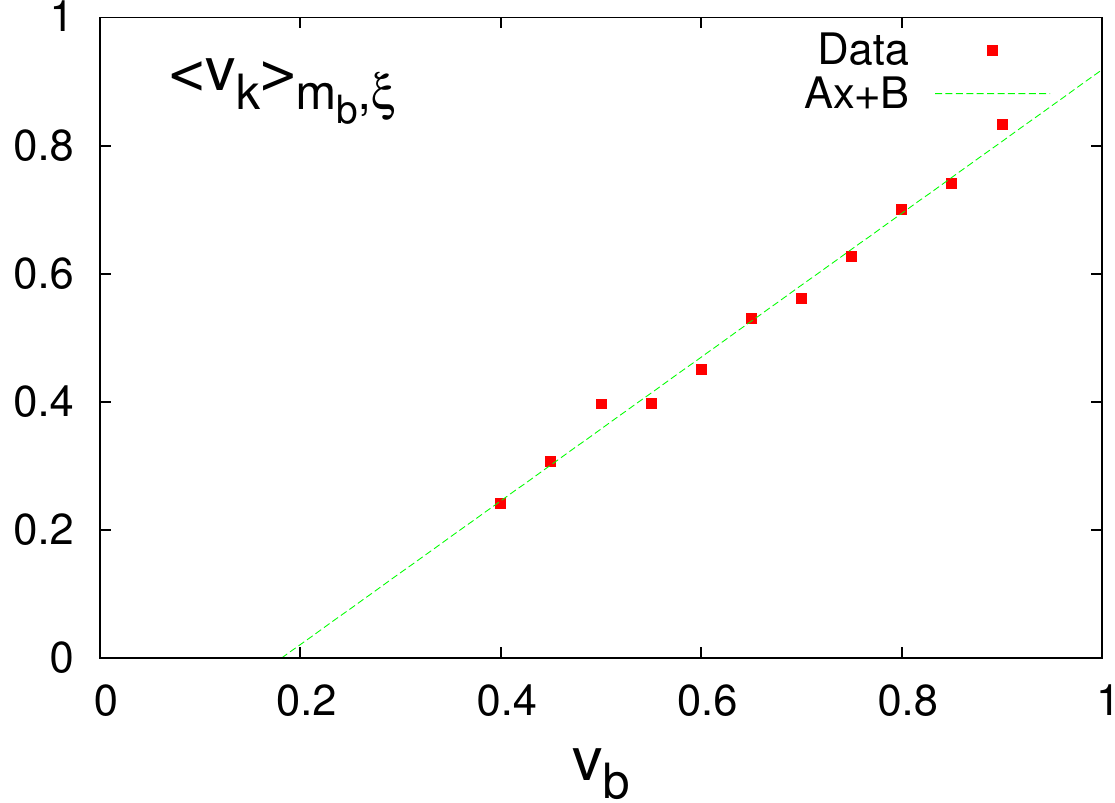}}
\subfloat[$\langle v_k\rangle_{v_b,m_b}$ vs.
$\xi$]{\includegraphics[width=0.33\linewidth]{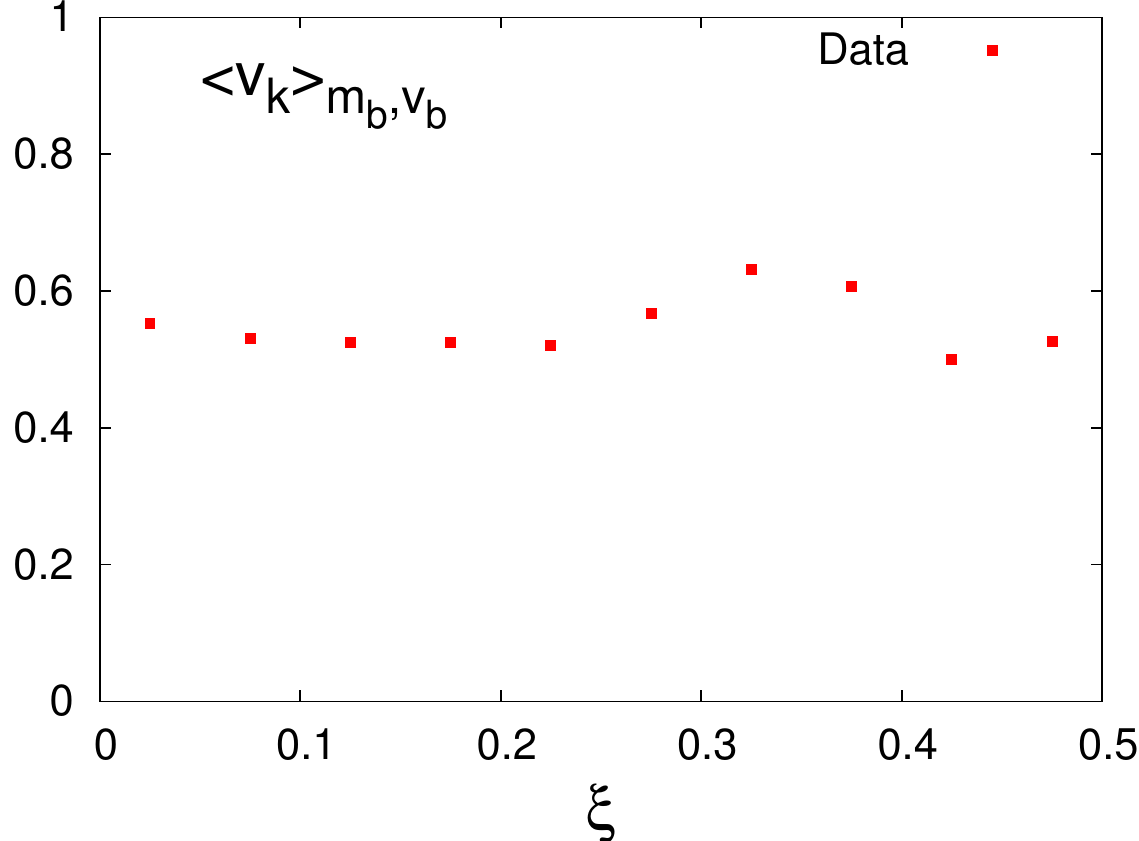}}
\caption{\label{fig:vkmean} Mean outgoing kink velocity as a function of initial conditions. In
each of these plots, two of the three initial parameters,  $m_b,v_k,\xi$ are summed over.}
\end{center}
\end{figure}
\end{widetext}

The trends in Fig.~\ref{fig:vkvsparams} can be quantified, as in Fig.~\ref{fig:vkmean}, where
we show how the mean value of $v_k$ depends on $m_b$, $v_b$, and $\xi$.
Linear fits give
\begin{equation}
\label{eq:kvvm}
\langle v_k \rangle_{v_b,\xi}=( 0.56 \pm 0.04 )m_b+( -0.34 \pm 0.05 )
\end{equation}

\begin{equation}
\langle v_k \rangle_{m_b,\xi}=( 1.12 \pm 0.04 )v_b+( -0.20 \pm 0.03)
\end{equation}

where the subscripts refer to the parameters over which the data is accumulated. If we choose parameters such that the mean $v_k$ is very
small, it implies that any kink-antikink pairs that are produced
will re-annihilate. Extrapolation from the above fits then suggests
that kink production will be suppressed for $m_b < 0.6\pm0.1$ and
$v_b < 0.18\pm0.03$.
As a function of $\xi$ we see that $v_k$ doesn't seem to vary dramatically, but we do note a slight increase in $v_k$ for $\xi=0.25-0.35$.

\subsection{Chaotic structure}
\label{chaos}

\begin{figure}
\begin{center}
\includegraphics[width=4 in]{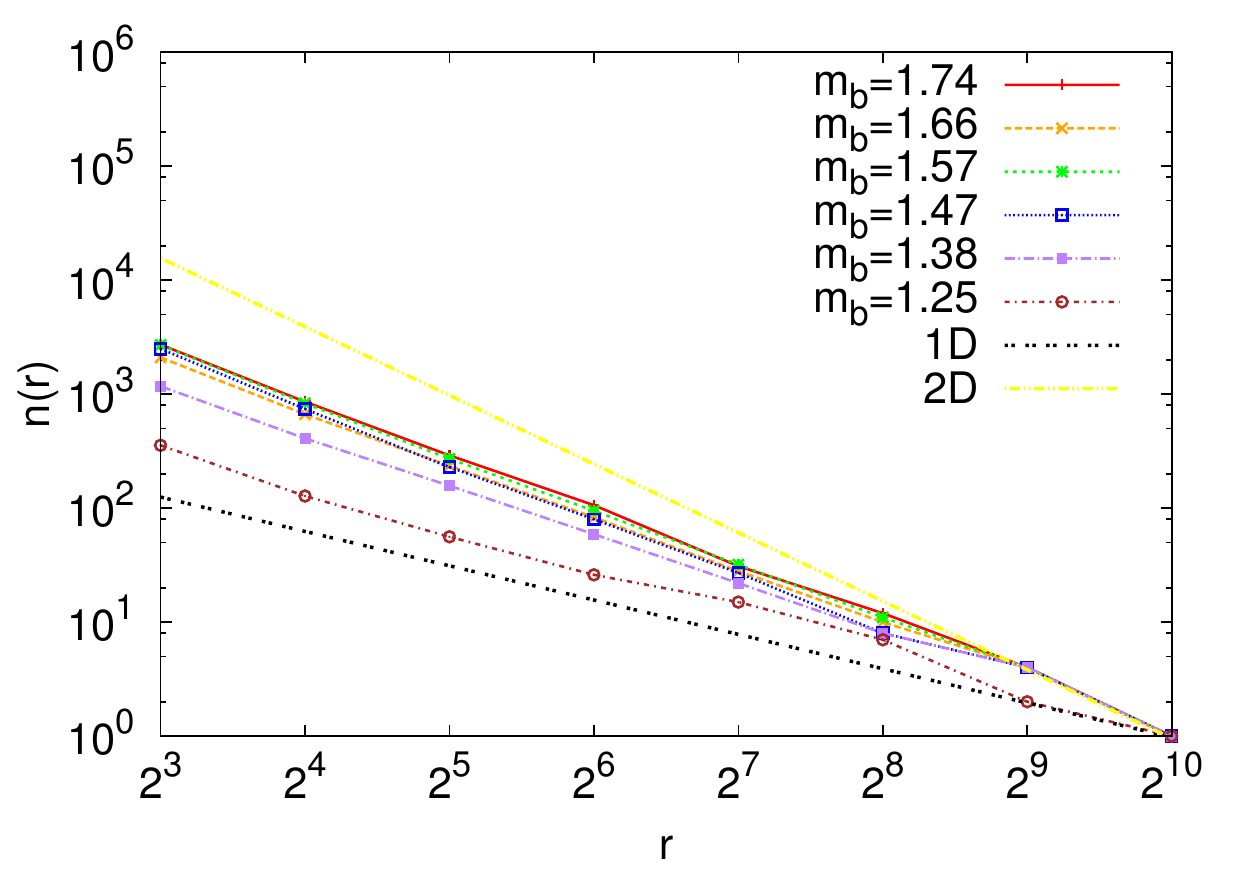}
\caption{
\label{f:boxs}
Number of boxes required to bound the area of successful kink production 
as a function of the box side length for various breather masses.  
$r$ is given in units of pixels.  The plot begins at $r=2^3$ because a point in our data corresponds to a $5\times5$ pixel box.  
}
\end{center}
\end{figure}

One way to quantify the chaotic nature of conditions is to consider the 
fractal dimension of plots of the outgoing kink velocities, $v_k$, as a 
function of the initial conditions as shown in Fig.~\ref{fig:kv}.  
Following the ideas of \cite{Romanczukiewicz:2010eg}, we
investigate a specific type of fractal dimension, the Minkowski-Bouligand
dimension or the box-counting dimension. This dimension is defined by
\begin{equation}
\label{eq:dim}
 D_{\rm box}=\lim_{r\rightarrow0}\frac{\log n(r)}{\log 1/r}
\end{equation}
where $n(r)$ is the number of boxes of side length $r$ that are required to
cover the outline of the shape considered.  
For a shape lacking fractal properties in
2D ({\it e.g.} a circle) we get $D_{\rm box}=1$; while an area
filling shape ({\it e.g.} a disc) has $D_{\rm box}=2$.  
A fractal shape in 2D has $1<D_{\rm box}\leq2$.  While 
Eq.~(\ref{eq:dim}) is the formal definition of the
box-counting dimension, in practice it can be difficult to extract from data.
Instead, since we expect the scaling of the boxes with $r$ to be of the form
\begin{equation}
 k(1/r)^{D_L}=n(r)
\end{equation}
where $D_L$ is called the local dimension and has a weak dependence on the box
size for small $r$.  Rearranging this equation we see that
\begin{equation}
 -D_L\log(r)+\log(k)=\log(n(r))
\end{equation}
So the local dimension will be given by
\begin{equation}
 D_L=-\frac{d\log(n(r))}{d \log(r)}
\end{equation}
and should be approximately $D_{\rm box}$ for small $r$. In Fig.~\ref{f:boxs} 
we plot the number of boxes needed to bound the area in Fig.~\ref{fig:kv} 
as a function of the side length.  We have found that for increasing $m_b$ 
there is a marked deviation from $D_{\rm box}=1$.

We compute $D_L$ at each box size by taking 2nd order central finite 
differences and, for averaging over all $r$ values, we arrive at a 
good estimate of box counting dimension, $D_{\rm box}$. The results for $D_{\rm box}$ 
are shown in Table \ref{tab:fd} for several $m_b$.  

\begin{table}
\caption{
\label{tab:fd}
Mean value and standard deviation of the mean for the box counting fractal 
dimension, $D_{\rm box}$, as a function of breather mass, $m_b$.  We note that this is in general smaller than a similar value ($1.770\pm0.011$) for the $\lambda \phi^4$ model found in \cite{Romanczukiewicz:2010eg} 
} 
\begin{tabular}{l|c} 
 \hline\hline 
 $m_b$ & $D_{box}$\\ 
 \hline 
1.74& 1.58 $\pm$ 0.03\\ 
1.66& 1.54 $\pm$ 0.04\\ 
1.57& 1.60 $\pm$ 0.04\\ 
1.47& 1.62 $\pm$ 0.06\\ 
1.38& 1.42 $\pm$ 0.05\\ 
1.25& 1.23 $\pm$ 0.10\\ 
\hline 
 \end{tabular}
\end{table}

\subsection{Dependence on Breather Mass}
\label{sec:mass}
\begin{figure}
\begin{center}
\subfloat{\includegraphics[width=0.5\linewidth]{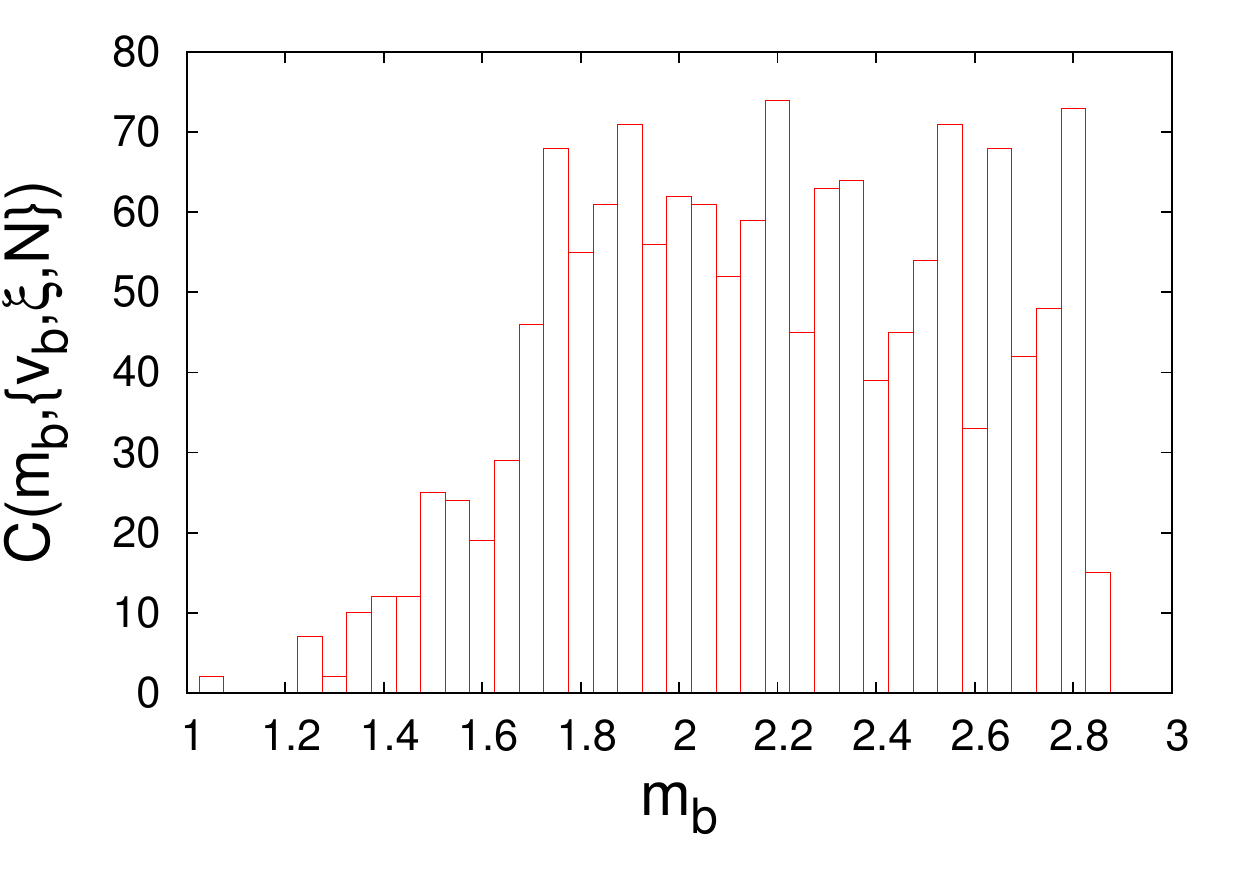}}
\subfloat{\includegraphics[width=0.5\linewidth]{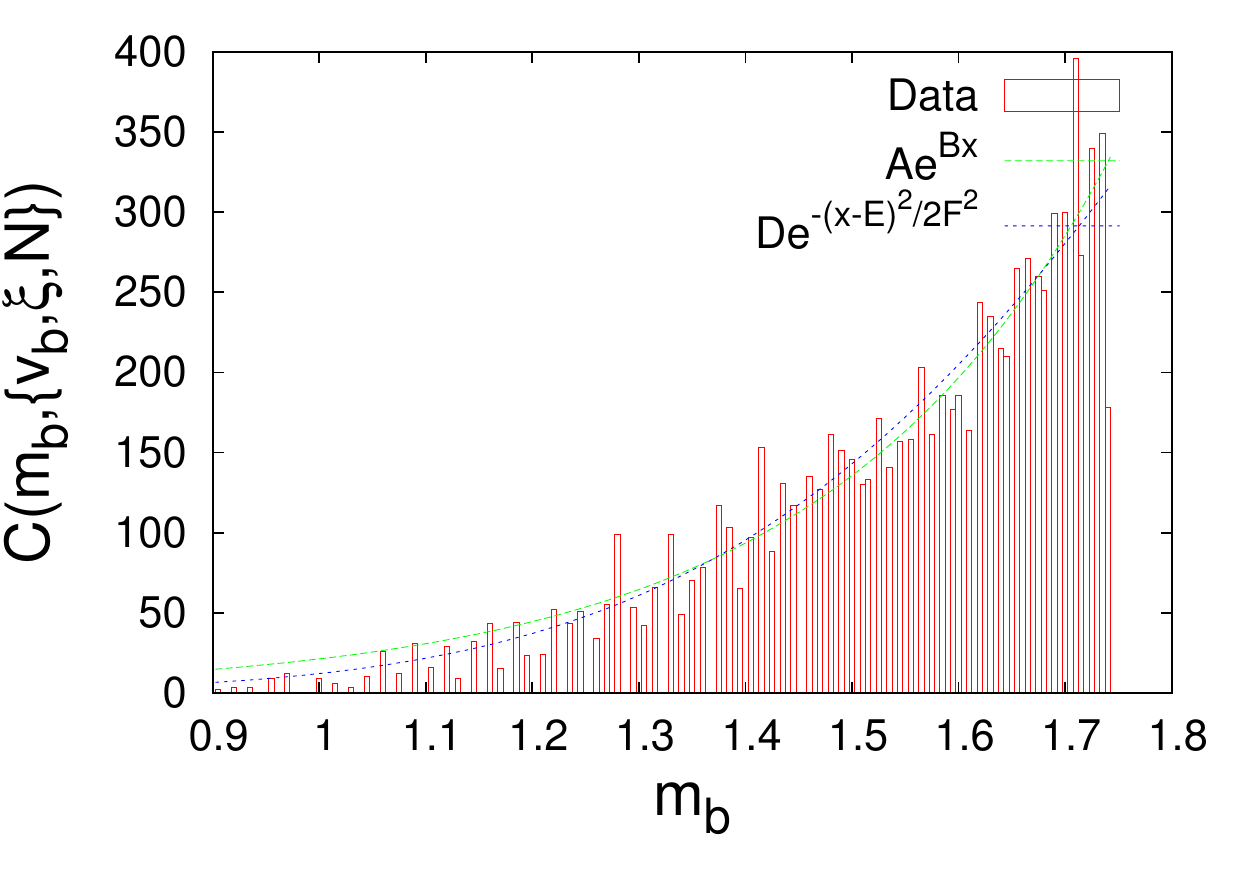}}
\caption{
\label{f:CvF}
Left: Accumulated number of successful events as a function of $m_b$ over the range $m_b \in (1,3)$ for all $v_b,\xi$ and $N \in (2,15)$. Right: A denser search over the range $m_b \in (0.9,1.8)$ for all $v_b,\xi$ and $N=4$}
\end{center}
\end{figure}

We expect that it is more difficult to produce kinks with low mass breathers
than it is with high mass breathers. This expectation is seen to be correct in 
Fig.~\ref{f:CvF}. We see that the success rate fluctuates around some value
for $m_b > 2$, where we note that the kink energy is $E_k=2$. While for
$m_b < 2$, the success rate decreases exponentially with decreasing $m_b$.
In the right panel of Fig.~\ref{f:CvF}, we searched the $m_b < 2$ region
more densely and fit the drop off with exponential and Gaussian profiles, 
with the Gaussian being marginally better in the low mass region. 

Focusing on the low mass region of Fig. \ref{f:CvF}, we see that the lowest mass at which 
soliton production is achieved is $m_b=0.9$.  Due to the exponential fall-off at small
breather mass, it would be very difficult to produce kinks using breathers of yet
lower mass and a random choice for the other parameters.

\subsection{Dependence on Incoming Breather Velocity}
\label{sec:vel}

\begin{figure}
\begin{center}
\includegraphics[width=4 in]{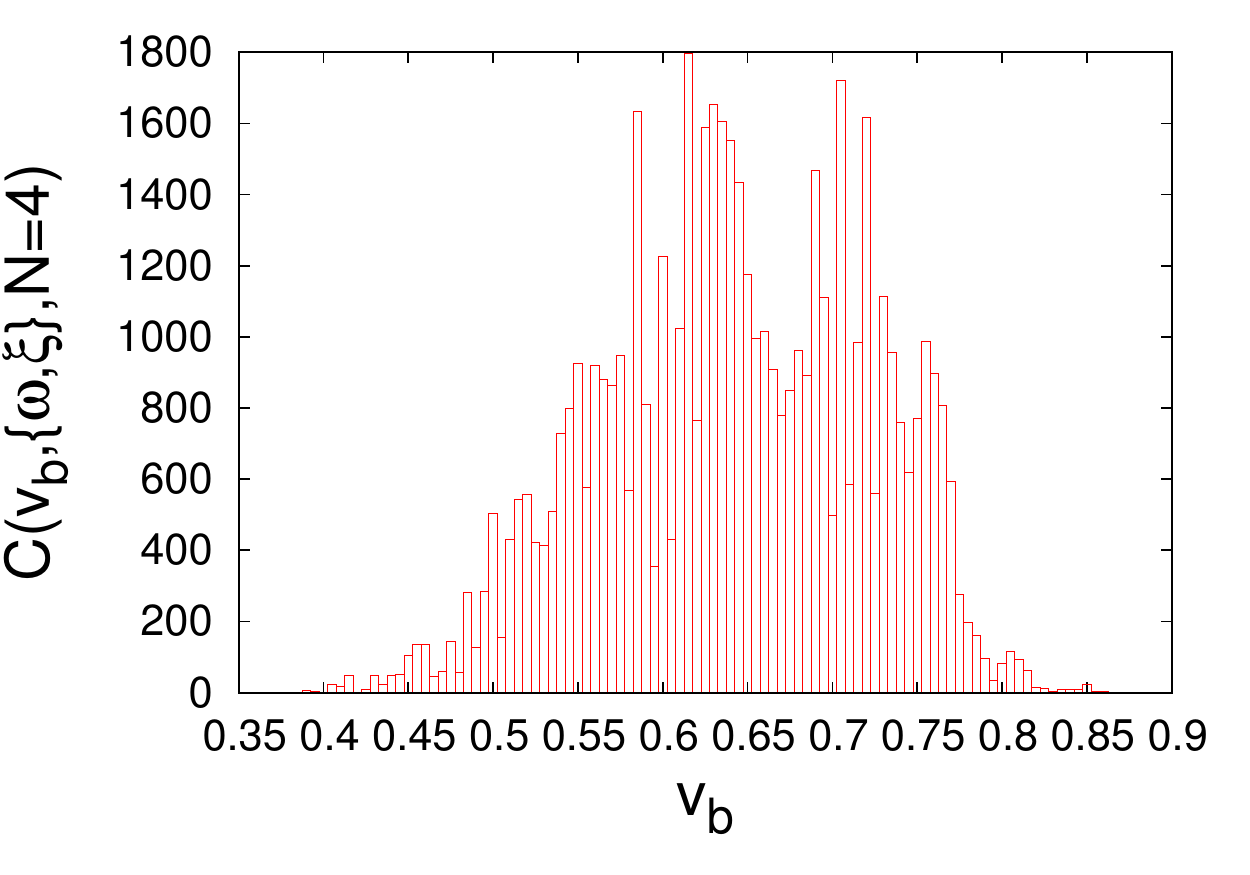}
\caption{\label{fig:cvv}Accumulated number of successful events as a function of $v_b$ 
over the range $m_b \in (0.9,1.8)$ for all $m_b,\xi$ and $N=4$ }
\end{center}
\end{figure}

Another important parameter for soliton production is the initial breather 
velocity.  From Fig.~\ref{fig:kv} we see that kink production occurs when the incoming
breather velocities lie in certain bands. These bands are insensitive to the breather
mass and survive even after we sum over all $m_b$ and $\xi$ as
seen in Fig.~\ref{fig:cvv} where distinct peaks are present.

\begin{figure}
\begin{center}
\includegraphics[width=4 in]{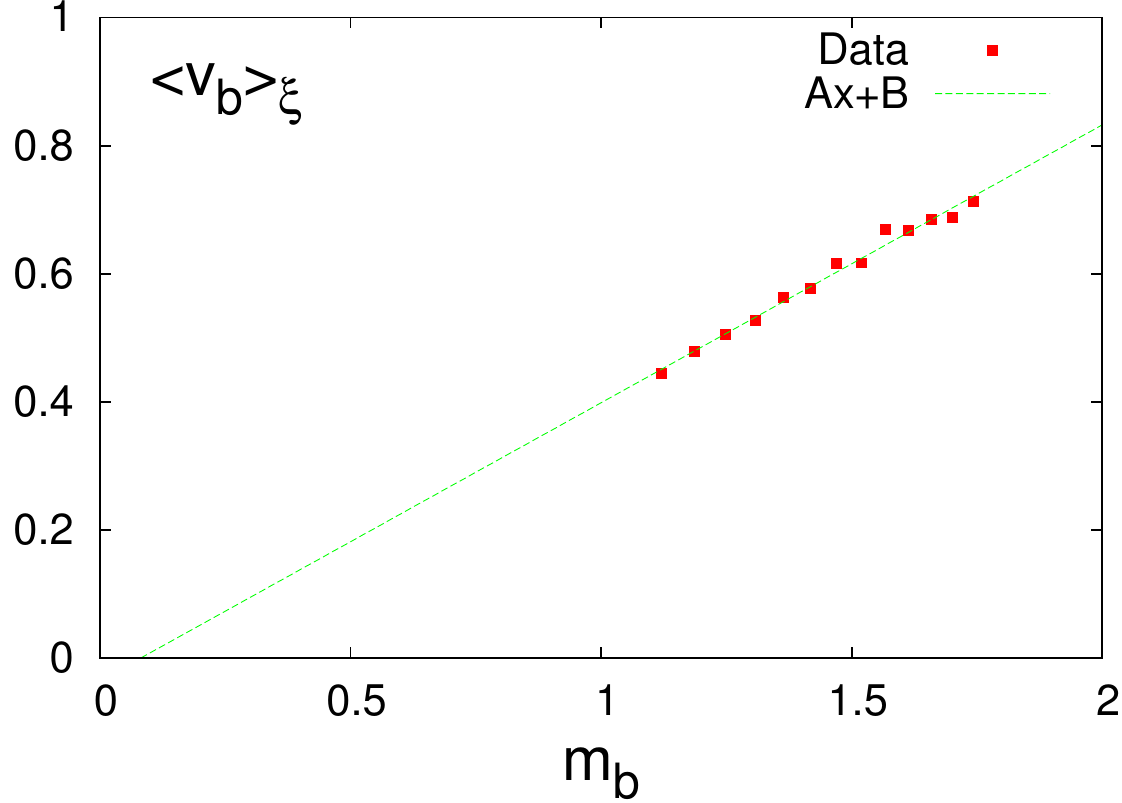}
\caption{\label{fig:vbmb}
$\langle {v}_b \rangle_{\xi}$ vs. $m_b$ for $N=4$.  From the fit,
we see that at $m_b=0.09\pm0.05$, $\langle{v}_b \rangle_\xi \to 0$ which suggests that kink production with such low mass breathers is highly suppressed.
}
\end{center}
\end{figure}

From our simulations, for a fixed breather mass, it is possible to calculate a mean breather 
velocity, $\langle {v}_b \rangle_{\xi}$,  at which kink production is most likely.
This value is found by accumulating the data from successful kink production over all twists 
and then finding the average breather velocity.  The data is plotted in Fig.~\ref{fig:vbmb}.
The dependence of $\langle v_b \rangle_\xi$ on $m_b$ is found to be linear in the region of $1.2<m_b<2$ with a fit
\begin{equation}
\langle v_b \rangle_{\xi}=( 0.43 \pm 0.02 )m_b+( -0.04 \pm 0.02 )
\end{equation}

Extrapolating to low mass, we obtain a mass cutoff at $m_{b}=0.09\pm0.05$. This bound
is much lower, hence weaker, than the bound of $m_b=0.6$ obtained by requiring that the 
outgoing kinks have a non-zero velocity (see above Eq.~\ref{eq:kvvm}).

We therefore see that for breathers of lower mass to produce kinks, the velocity at which they 
must be collided is lower. But with lower breather velocity, the outgoing kink velocity also
decreases, and at some point the kinks cannot escape and instead they re-annihilate. 
These results together indicate that kink production will be highly suppressed
in low $m_b$ regions in the ($m_b,v_b$) parameter space.

\subsection{Dependence on Twist}
\label{sec:twist}

\begin{figure}
\begin{center}
\includegraphics[width=4 in]{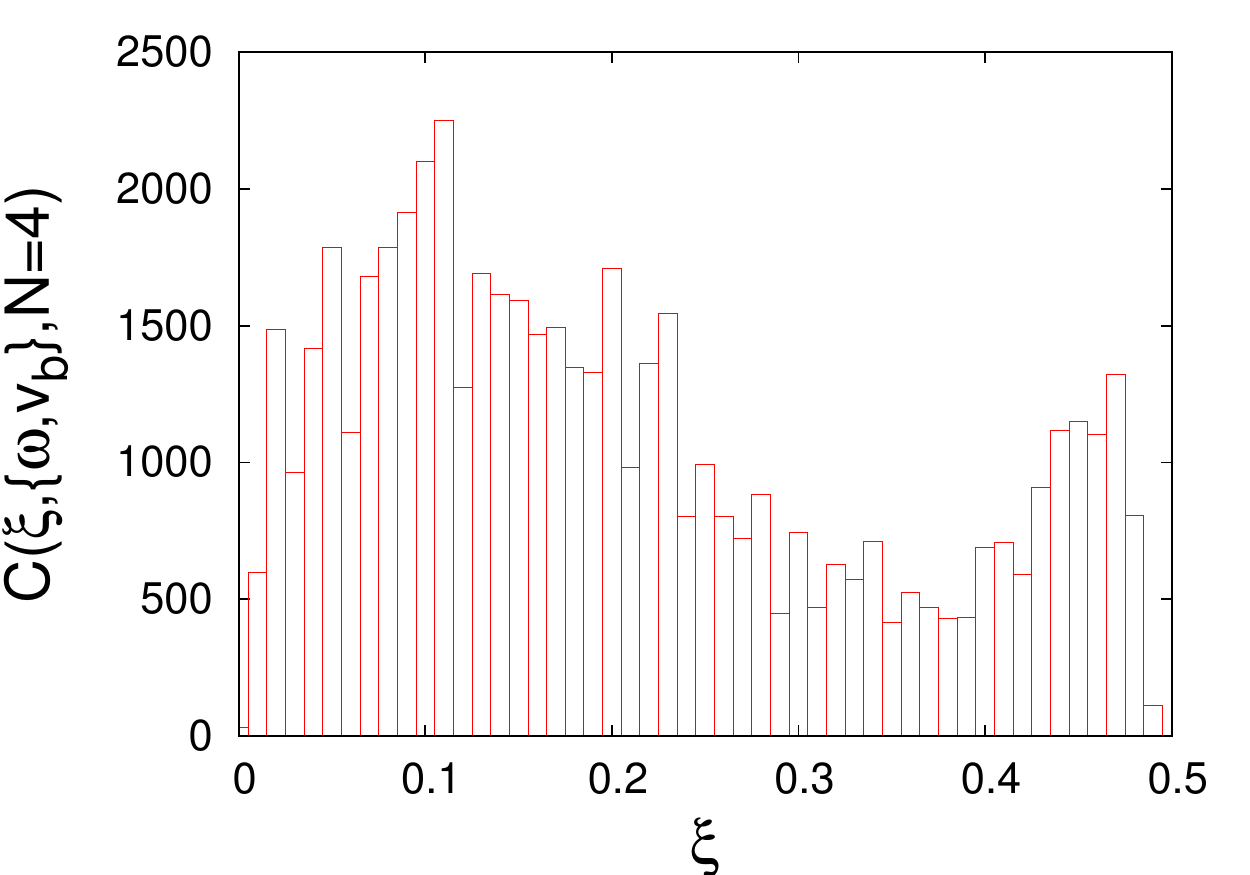}
\caption{
\label{fig:cvt}
Number of kinks produced vs $\xi$, summed over $v_b$, $m_b$ and for $N=4$.  Notice that for $\xi=0.25-0.4$ there is a distinct drop in production } 
\end{center}
\end{figure}

An intuitive understanding of how the success of production should depend upon $\xi$ is not
obvious, but a few simple characteristics are expected.  For $\xi=0,0.5$, the dynamics is
the same as in the sine-Gordon model and we therefore expect there to be no kink production 
at these points.  From Fig.~\ref{fig:cvt}, we see that this is in fact true.  In the same figure, we 
see that even for small deviations from the sine-Gordon model, the production rate is quite large.

In previous sections, it has been seen that in a region located around $\xi=0.25-0.40$ various interesting features occur.  We have shown that the likelihood of production is decreased in this region (Fig.~\ref{fig:kv}) and that the average outgoing kink velocity is slightly increased 
(Fig.~\ref{fig:vkmean}).  
From Fig.~\ref{fig:cvt} we see that the production count decrease in this region is over $40\%$, indicating some interesting physics must be occurring.

To understand the strange effects observed at $\xi=0.25-0.40$ on twist, we investigated the case of two breathers whose centers of energy are initially at rest.  Previous calculations have found that the force between two breathers in the sine-Gordon model is to first order \cite{PhysRevE.80.036603}
\begin{equation}
\label{eq:sgforce}
 F=\mp16\eta^4\omega^2e^{-2\eta\omega |L|}
\end{equation} 
where the negative (positive) sign indicates attraction (repulsion) for in-phase (out-of-phase) breathers, and $L$ is half the distance between the breathers.
Since the phase of the breathers also corresponds to $\xi=0$ for in-phase breathers and
$\xi=0.5$ for out-of-phase breathers, we expect that the force should be attractive for low 
twist, and repulsive for high twist in the \o3z breather case.  In \cite{Vachaspati:2011ad}, 
the time delay was numerically investigated for the \o3z model and it was found that for 
$\xi=0.2-0.4$ there was dramatic increase in time delay over the $\xi=0,0.5$ cases.  Both
of these works seem to indicate that at intermediate values of $\xi$ there is something
different about the breather interactions.

\begin{figure}
 \begin{center}
 \subfloat[$m_b=1.74$]{\includegraphics[width=0.5\linewidth]{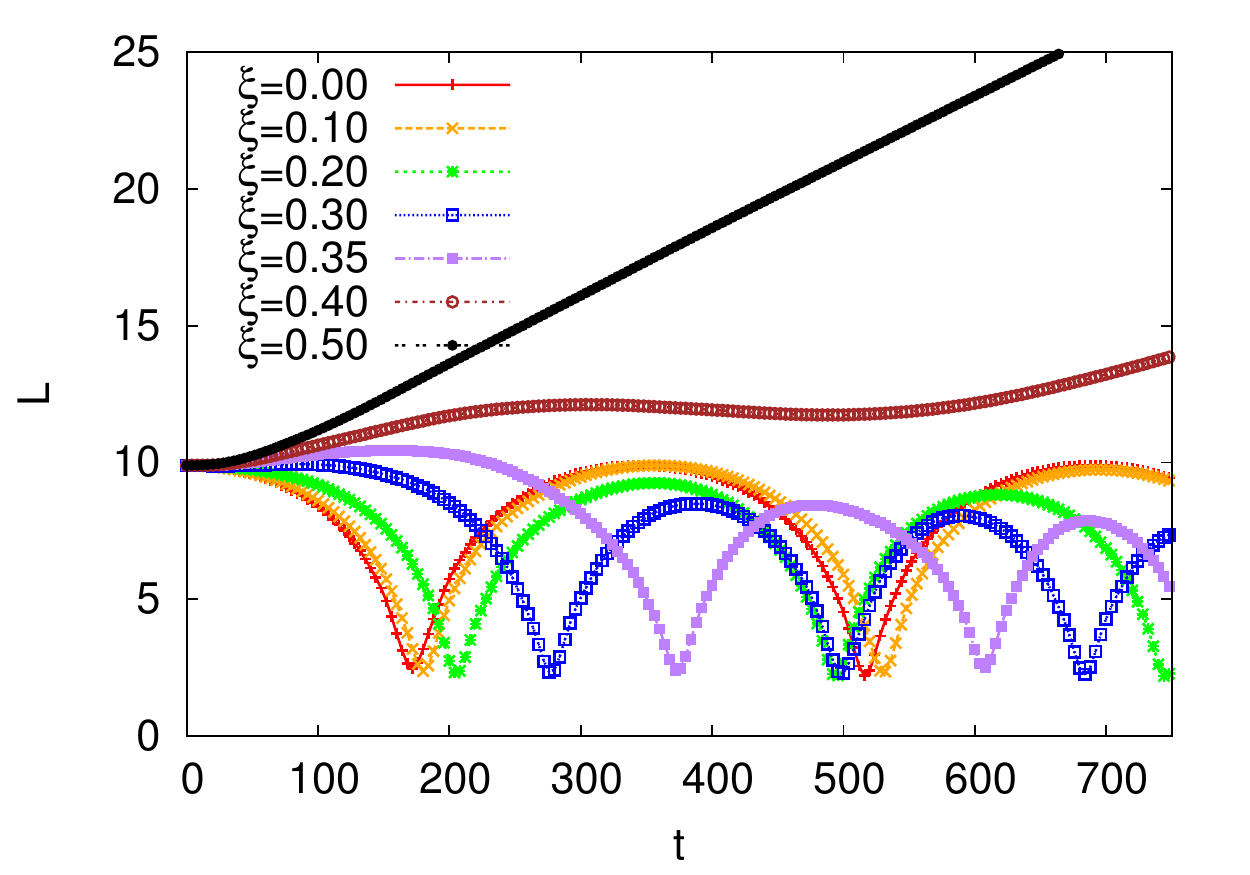}}
 \subfloat[$m_b=1.25$]{\includegraphics[width=0.5\linewidth]{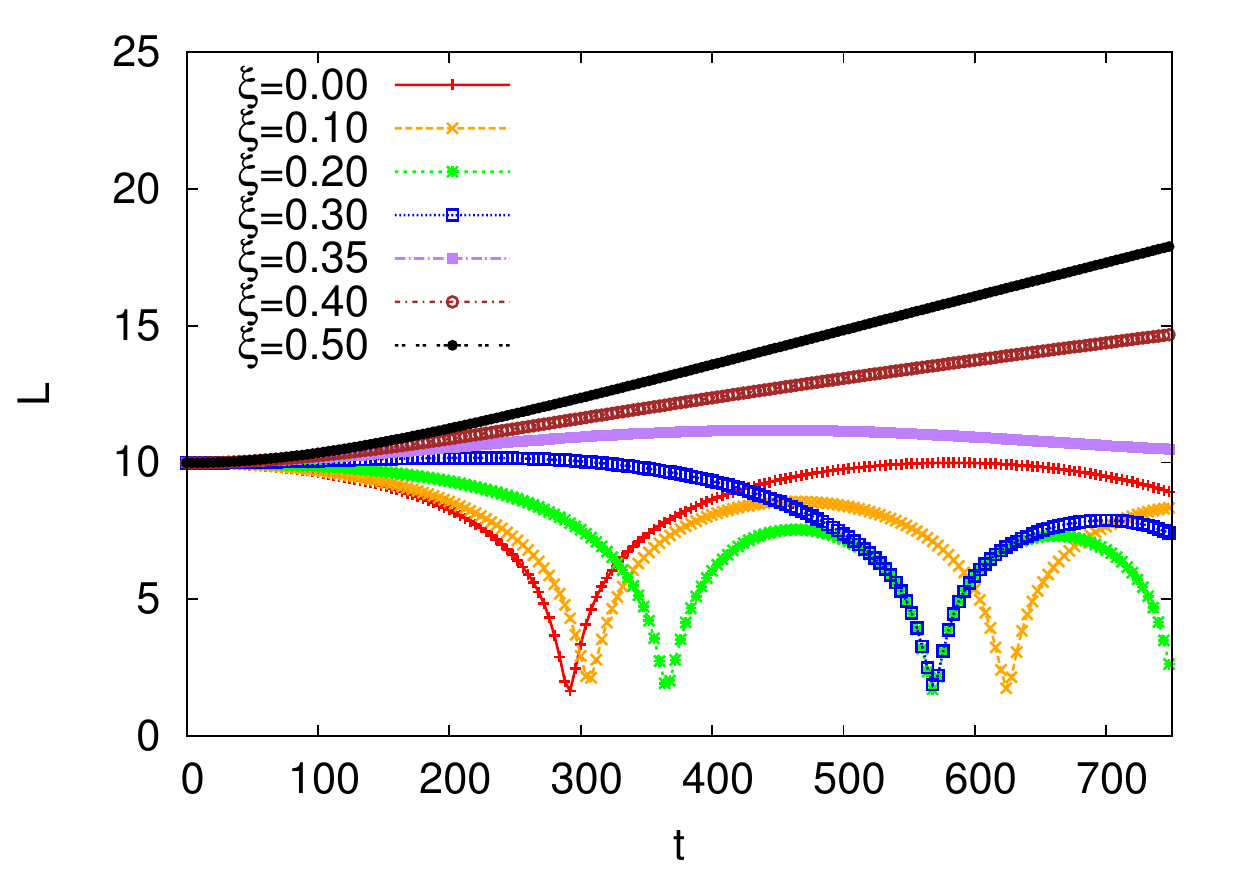}}
\caption[$x(t)$ for various $\xi$]{\label{fig:sepvtwist}Time evolution of the
separation distance between two breather center of energies for
$\xi=0,0.10,0.20,0.30,0.35,0.40,0.50$ for $m_b=1.74,1.25$.  Below a critical
value of $\xi\approx0.35-0.40$ the force is attractive and the breathers
oscillate about a shared center; above the critical value, the breathers repel
and move apart.}
\end{center}
\end{figure}

We have studied the dynamics of two initially static breathers for two different values of
$m_b$ and for various values of $\xi$. The results in Fig.~\ref{fig:sepvtwist} show that there
is a critical value of twist at which the force changes from being attractive to being
repulsive. The critical value depends on $m_b$ and is larger for lower mass breathers.
This suggests that the observed drop in kink production around $\xi \approx 0.3$ might 
be correlated with the lack of interaction between breathers at the critical twist.

\subsection{Dependence on Number of Breathers}
\label{sec:num}
\begin{figure}
\begin{center}
\includegraphics[width=4 in]{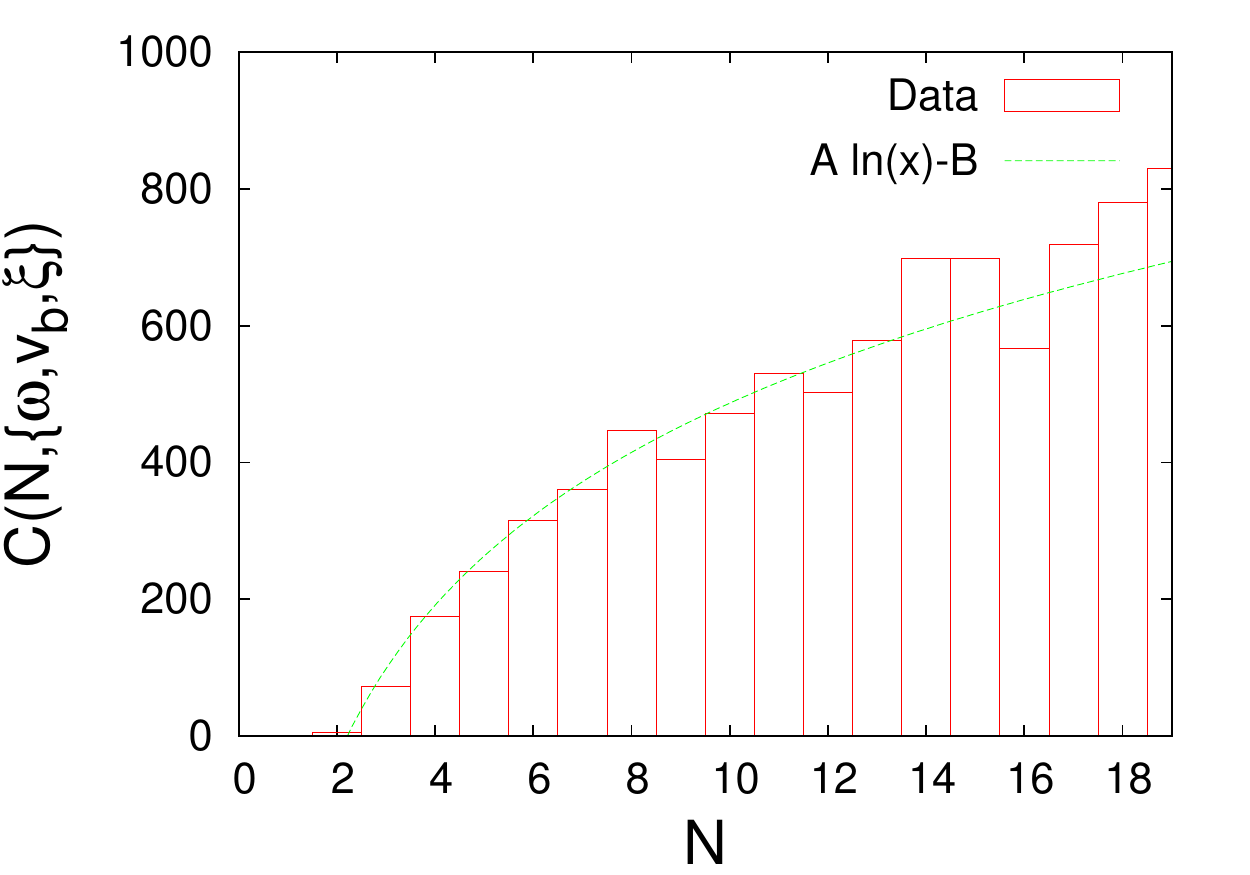}
\caption{
\label{fig:cvn}
Number of kinks produced vs. $N$, summed over $v_b$, $m_b$ and $\xi$.  The likelihood of kink production increases, but slows with increasing $N$.  A good fit was found to be 
$(330\pm30)\ln(N)-(280\pm70)$.
}
\end{center}
\end{figure}

As one might expect, increasing the number of breathers in the initial train of breathers 
increases the chances of kink production. This data is shown in Fig.~\ref{fig:cvn}.  The shown
fit corresponds to logarithmic growth of successful kink production events with the 
number of breathers. So the gain in kink production depends weakly on the number
of breathers. This agrees with our earlier discussion in the context of the left panel
of Fig.~\ref{fig:NvE}, where we suggested that it is the energy per breather that is important
for kink production and not so much the total energy in the train of breathers.

To further understand the effects of $N$, we investigated a set of initial conditions where for 
$N=4$ we know solitons are produced, and then we varied $N$.  We found, as seen in 
Fig.~\ref{fig:focus}, that the solitons were produced in the collision of the second breathers 
in each train before the third and fourth had had a chance to collide. This result is at first troubling because for the cases of $N=1,2,3$ it is found that no kinks are produced.  So the fourth
breather in the train is critical to kink production. The exact role that the fourth breather
plays is not clear to us though some possibilities come to mind. The fourth breather
may influence the forward breathers in the train prior to the collision and change
some of their characteristics, and this is what enables kink production. Another possibility 
is that the collisions of the number two
breathers produce a proto-kink pair which requires additional momentum transfer via
the fourth breather to grow into a kink-antikink pair.

We speculate that the diminishing return on success of additional breathers arises from the limitation imposed on interaction between breathers based on their initial spacing in the trains.  Although the numbers of breathers increases, the distance between the beginning and end of the train also increases.  Since the force between breathers is generically exponential with distance (i.e. Eq.~\ref{eq:forceg}), breathers that are sufficiently far away have little effect on each other.

\subsection{Other Initial Conditions}
\label{sec:initialvel}

In Sec.~\ref{sec:in}, it was discussed that there is a freedom in choosing $\dot{\phi}(t=0)$.
In the previous discussions we have only studied the simplest case of $\dot{\phi}(t=0)=0$.  Here we consider two other possible set of initial conditions.  The first type is the ``co-spinning'' initial conditions
\begin{equation}
 \dot{\phi}(t=0,x)=v_{\phi}
\end{equation}
where $v_{\phi}$ is some constant velocity and all the breather trains rotate have the same
velocity in the $\phi$ variable.  Additionally, we considered ``counter-spinning'' initial conditions
\begin{equation}
 \dot{\phi}(t=0,x)=v_{\phi}\tanh(x/w)
\end{equation}
where the left-moving and right-moving breather trains have opposite velocity in the
$\phi$ direction. These choices both introduce another free parameter into the initial conditions, namely, the initial $\phi$ velocity $v_{\phi}$.  On inclusion of a $\dot{\phi}$ term into the equations of motion, the sine-Gordon breathers are no longer exact solutions, but from our simulations they are found to still be long-lived.   

Spinning initial conditions may also be viewed as charged initial conditions, following the 
remark below Eq.~(\ref{eom}). Co-spinning initial conditions correspond to like charges on
the incoming breather trains; counter-spinning initial conditions correspond to opposite charges on
the breather trains.

\begin{widetext}
\begin{figure}
\begin{center}
\includegraphics[width=\linewidth]{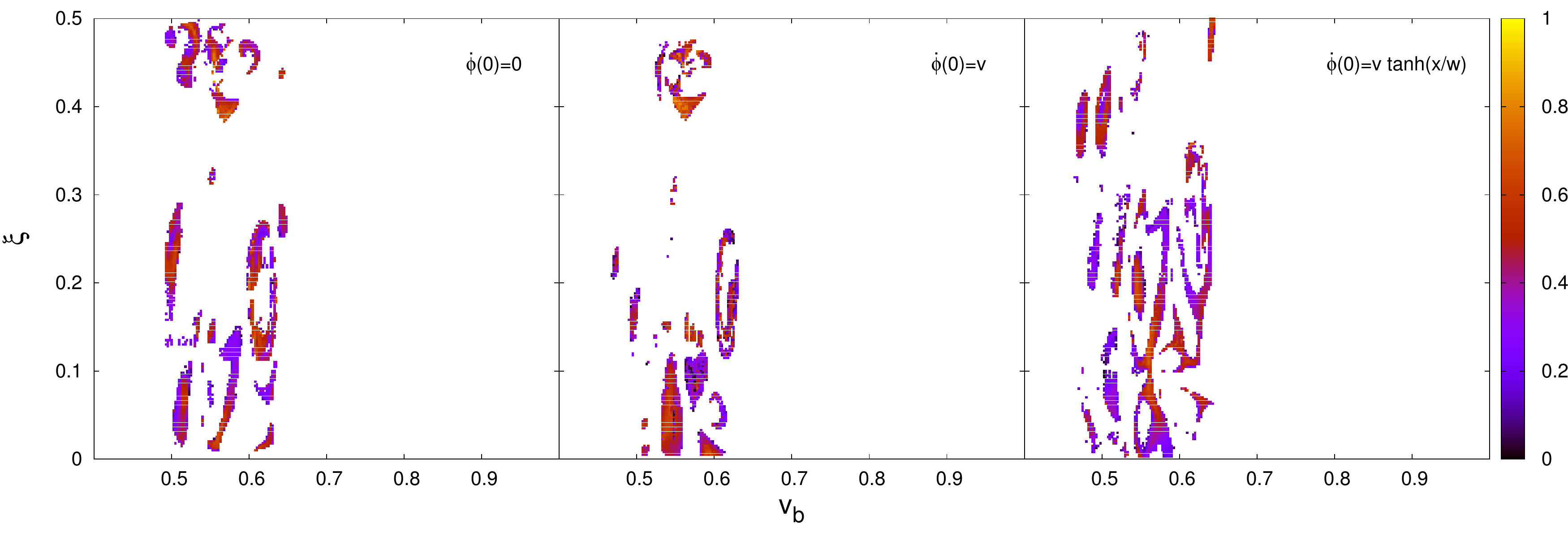}
\caption{\label{fig:kv_spin} Velocity of outgoing kinks (denoted by color) as a function 
of incoming breather velocity and twist for the non-spinning (left panel), co-spinning (center
panel), and counter-spinning (right panel) initial conditions for $m_b=1.36$, $v=0.1$, and $N=4$.}
\end{center}
\end{figure} 
\end{widetext}

We have run our simulations for the co-spinning and counter-spinning initial conditions with 
$v_{\phi}=0.1$, $m_b=1.36$, and $N=4$ and some of the results are shown in
Fig.~\ref{fig:kv_spin}. The co-spinning initial conditions have only 
$80\%$ of the number of successful kink production events as the non-spinning initial
conditions for the same phase space search. Furthermore, the distributions in breather velocity 
and twist are shifted to lower values and have smaller spreads.  The mean outgoing kink velocity 
is also decreased.  Putting these trends together, we conjecture that co-spinning initial conditions suppress kink production, but additional study should be undertaken to confirm this.

Counter-spinning initial conditions appear to have exactly the opposite effects.  The spread in 
breather velocities and twist increased, and the final kink velocities also increased.  For the same parameter space search, counter-spinning runs were more successful than
non-spinning runs by a factor of 1.4, suggesting that such initial conditions may be worth
exploring further.

\section{Conclusions}
\label{sec:con}

We have numerically explored a wide range of scattering initial conditions in the \o3z model 
that can lead to the production of a kink-antikink pair. Our initial state consists of two 
oppositely moving trains of breather solutions. There are several general features that
we have observed that we now summarize: (i) the region in parameter space that
leads to kink production has fractal structure, (ii) smaller breathers need to be scattered
at smaller velocities, (iii) when kinks are produced, their outgoing velocities increase
in proportion to the incoming breather train velocity, (iv) twist is essential for 
kink production in this model but the outcomes are not strongly sensitive to the exact
value that we choose, and (v) the force between breathers vanishes for a certain
value of the twist.
Putting together points (ii) and (iii) we conclude that small breather
velocities are necessary for building kinks, while large breather velocities help to
separate them. Hence there is tension in the requirements for successful kink production
and we can expect that the process will be highly suppressed when the breather
mass is small compared to the kink mass.

These pessimistic conclusions are somewhat offset by our finding that counter-spinning
initial conditions can enhance kink production. Further exploration of such initial
conditions may lead to better understanding of when kinks can be (more easily) produced.

Finally we observe that, since the sine-Gordon model is embedded 
inside the \o3z model, and soliton operators have been constructed in the sine-Gordon
model \cite{Mandelstam:1975hb}, it is possible that similar operators can be found
in the quantum \o3z model. Then it is conceivable that kink production 
can be studied in the \o3z model by quantum field theory methods or on a lattice.

 \begin{acknowledgments}
The numerical work was done on the Saguaro cluster at the ASU Advanced Computing Center.
This work was supported by the DOE at ASU.
 \end{acknowledgments}

\bibliographystyle{apsrev4-1}
\bibliography{mybib}

\begin{thebibliography}{10}%
\makeatletter
\providecommand \@ifxundefined [1]{%
 \ifx #1\undefined \expandafter \@firstoftwo
 \else \expandafter \@secondoftwo
\fi
}%
\providecommand \@ifnum [1]{%
 \ifnum #1\expandafter \@firstoftwo
 \else \expandafter \@secondoftwo
\fi
}%
\providecommand \enquote [1]{``#1''}%
\providecommand \bibnamefont  [1]{#1}%
\providecommand \bibfnamefont [1]{#1}%
\providecommand \citenamefont [1]{#1}%
\providecommand\href[0]{\@sanitize\@href}%
\providecommand\@href[1]{\endgroup\@@startlink{#1}\endgroup\@@href}%
\providecommand\@@href[1]{#1\@@endlink}%
\providecommand \@sanitize [0]{\begingroup\catcode`\&12\catcode`\#12\relax}%
\@ifxundefined \pdfoutput {\@firstoftwo}{%
 \@ifnum{\z@=\pdfoutput}{\@firstoftwo}{\@secondoftwo}%
}{%
 \providecommand\@@startlink[1]{\leavevmode\special{html:<a href="#1">}}%
 \providecommand\@@endlink[0]{\special{html:</a>}}%
}{%
 \providecommand\@@startlink[1]{%
  \leavevmode
  \pdfstartlink
   attr{/Border[0 0 1 ]/H/I/C[0 1 1]}%
   user{/Subtype/Link/A<</Type/Action/S/URI/URI(#1)>>}%
  \relax
 }%
 \providecommand\@@endlink[0]{\pdfendlink}%
}%
\providecommand \url  [0]{\begingroup\@sanitize \@url }%
\providecommand \@url [1]{\endgroup\@href {#1}{\urlprefix}}%
\providecommand \urlprefix [0]{URL }%
\providecommand \Eprint[0]{\href }%
\@ifxundefined \urlstyle {%
  \providecommand \doi [1]{doi:\discretionary{}{}{}#1}%
}{%
  \providecommand \doi [0]{doi:\discretionary{}{}{}\begingroup
  \urlstyle{rm}\Url }%
}%
\providecommand \doibase [0]{http://dx.doi.org/}%
\providecommand \Doi[1]{\href{\doibase#1}}%
\providecommand \bibAnnote [3]{%
  \BibitemShut{#1}%
  \begin{quotation}\noindent
    \textsc{Key:}\ #2\\\textsc{Annotation:}\ #3%
  \end{quotation}%
}%
\providecommand \bibAnnoteFile [2]{%
  \IfFileExists{#2}{\bibAnnote {#1} {#2} {\input{#2}}}{}%
}%
\providecommand \typeout [0]{\immediate \write \m@ne }%
\providecommand \selectlanguage [0]{\@gobble}%
\providecommand \bibinfo [0]{\@secondoftwo}%
\providecommand \bibfield [0]{\@secondoftwo}%
\providecommand \translation [1]{[#1]}%
\providecommand \BibitemOpen[0]{}%
\providecommand \bibitemStop [0]{}%
\providecommand \bibitemNoStop [0]{.\EOS\space}%
\providecommand \EOS [0]{\spacefactor3000\relax}%
\providecommand \BibitemShut [1]{\csname bibitem#1\endcsname}%
\bibitem{Mottola:1988ff}%
  \BibitemOpen
  \bibfield{author}{%
  \bibinfo {author} {\bibfnamefont{E.}~\bibnamefont{Mottola}}\ and\ \bibinfo
  {author} {\bibfnamefont{A.}~\bibnamefont{Wipf}},\ }%
  \bibfield{journal}{%
  \Doi{10.1103/PhysRevD.39.588}{\bibinfo {journal} {Phys.Rev.}}\ }%
  \textbf{\bibinfo {volume} {D39}},\ \bibinfo {pages} {588} (\bibinfo {year}
  {1989})%
  \bibAnnoteFile{NoStop}{Mottola:1988ff}%
\bibitem{Ringwald:1989ee}%
  \BibitemOpen
  \bibfield{author}{%
  \bibinfo {author} {\bibfnamefont{A.}~\bibnamefont{Ringwald}},\ }%
  \bibfield{journal}{%
  \Doi{10.1016/0550-3213(90)90300-3}{\bibinfo {journal} {Nucl.Phys.}}\ }%
  \textbf{\bibinfo {volume} {B330}},\ \bibinfo {pages} {1} (\bibinfo {year}
  {1990})%
  \bibAnnoteFile{NoStop}{Ringwald:1989ee}%
\bibitem{Mattis:1991bj}%
  \BibitemOpen
  \bibfield{author}{%
  \bibinfo {author} {\bibfnamefont{M.~P.}\ \bibnamefont{Mattis}},\ }%
  \bibfield{journal}{%
  \Doi{10.1016/0370-1573(92)90033-V}{\bibinfo {journal} {Phys.Rept.}}\ }%
  \textbf{\bibinfo {volume} {214}},\ \bibinfo {pages} {159} (\bibinfo {year}
  {1992})%
  \bibAnnoteFile{NoStop}{Mattis:1991bj}%
\bibitem{Rebbi:1996zx}%
  \BibitemOpen
  \bibfield{author}{%
  \bibinfo {author} {\bibfnamefont{C.}~\bibnamefont{Rebbi}}\ and\ \bibinfo
  {author} {\bibfnamefont{J.}~\bibnamefont{Singleton},
  \bibfnamefont{Robert~L.}},\ }%
  \bibfield{journal}{%
  \Doi{10.1103/PhysRevD.54.1020}{\bibinfo {journal} {Phys.Rev.}}\ }%
  \textbf{\bibinfo {volume} {D54}},\ \bibinfo {pages} {1020} (\bibinfo {year}
  {1996}),\ \Eprint{http://arxiv.org/abs/hep-ph/9601260}{arXiv:hep-ph/9601260
  [hep-ph]}%
  \bibAnnoteFile{NoStop}{Rebbi:1996zx}%
\bibitem{Kuznetsov:1997az}%
  \BibitemOpen
  \bibfield{author}{%
  \bibinfo {author} {\bibfnamefont{A.}~\bibnamefont{Kuznetsov}}\ and\ \bibinfo
  {author} {\bibfnamefont{P.}~\bibnamefont{Tinyakov}},\ }%
  \bibfield{journal}{%
  \Doi{10.1103/PhysRevD.56.1156}{\bibinfo {journal} {Phys.Rev.}}\ }%
  \textbf{\bibinfo {volume} {D56}},\ \bibinfo {pages} {1156} (\bibinfo {year}
  {1997}),\ \Eprint{http://arxiv.org/abs/hep-ph/9703256}{arXiv:hep-ph/9703256
  [hep-ph]}%
  \bibAnnoteFile{NoStop}{Kuznetsov:1997az}%
\bibitem{Bezrukov:2003er}%
  \BibitemOpen
  \bibfield{author}{%
  \bibinfo {author} {\bibfnamefont{F.}~\bibnamefont{Bezrukov}}, \bibinfo
  {author} {\bibfnamefont{D.}~\bibnamefont{Levkov}}, \bibinfo {author}
  {\bibfnamefont{C.}~\bibnamefont{Rebbi}}, \bibinfo {author}
  {\bibfnamefont{V.}~\bibnamefont{Rubakov}},\ and\ \bibinfo {author}
  {\bibfnamefont{P.}~\bibnamefont{Tinyakov}},\ }%
  \bibfield{journal}{%
  \Doi{10.1103/PhysRevD.68.036005}{\bibinfo {journal} {Phys.Rev.}}\ }%
  \textbf{\bibinfo {volume} {D68}},\ \bibinfo {pages} {036005} (\bibinfo {year}
  {2003}),\ \Eprint{http://arxiv.org/abs/hep-ph/0304180}{arXiv:hep-ph/0304180
  [hep-ph]}%
  \bibAnnoteFile{NoStop}{Bezrukov:2003er}%
\bibitem{Levkov:2004tf}%
  \BibitemOpen
  \bibfield{author}{%
  \bibinfo {author} {\bibfnamefont{D.}~\bibnamefont{Levkov}}\ and\ \bibinfo
  {author} {\bibfnamefont{S.}~\bibnamefont{Sibiryakov}},\ }%
  \bibfield{journal}{%
  \Doi{10.1103/PhysRevD.71.025001}{\bibinfo {journal} {Phys.Rev.}}\ }%
  \textbf{\bibinfo {volume} {D71}},\ \bibinfo {pages} {025001} (\bibinfo {year}
  {2005}),\ \Eprint{http://arxiv.org/abs/hep-th/0410198}{arXiv:hep-th/0410198
  [hep-th]}%
  \bibAnnoteFile{NoStop}{Levkov:2004tf}%
\bibitem{Demidov:2011dk}%
  \BibitemOpen
  \bibfield{author}{%
  \bibinfo {author} {\bibfnamefont{S.}~\bibnamefont{Demidov}}\ and\ \bibinfo
  {author} {\bibfnamefont{D.}~\bibnamefont{Levkov}},\ }%
  \bibfield{journal}{%
  \Doi{10.1103/PhysRevLett.107.071601}{\bibinfo {journal} {Phys.Rev.Lett.}}\ }%
  \textbf{\bibinfo {volume} {107}},\ \bibinfo {pages} {071601} (\bibinfo {year}
  {2011}),\ \Eprint{http://arxiv.org/abs/1103.0013}{arXiv:1103.0013 [hep-th]}%
  \bibAnnoteFile{NoStop}{Demidov:2011dk}%
\bibitem{Vachaspati:2011ad}%
  \BibitemOpen
  \bibfield{author}{%
  \bibinfo {author} {\bibfnamefont{T.}~\bibnamefont{Vachaspati}},\ }%
  \bibfield{journal}{%
  \Doi{10.1103/PhysRevD.84.125003}{\bibinfo {journal} {Phys.Rev.}}\ }%
  \textbf{\bibinfo {volume} {D84}},\ \bibinfo {pages} {125003} (\bibinfo {year}
  {2011}),\ \Eprint{http://arxiv.org/abs/1109.1065}{arXiv:1109.1065 [hep-th]}%
  \bibAnnoteFile{NoStop}{Vachaspati:2011ad}%
\bibitem{Dutta:2008jt}%
  \BibitemOpen
  \bibfield{author}{%
  \bibinfo {author} {\bibfnamefont{S.}~\bibnamefont{Dutta}}, \bibinfo {author}
  {\bibfnamefont{D.~A.}\ \bibnamefont{Steer}},\ and\ \bibinfo {author}
  {\bibfnamefont{T.}~\bibnamefont{Vachaspati}},\ }%
  \bibfield{journal}{%
  \Doi{10.1103/PhysRevLett.101.121601}{\bibinfo {journal} {Phys.Rev.Lett.}}\ }%
  \textbf{\bibinfo {volume} {101}},\ \bibinfo {pages} {121601} (\bibinfo {year}
  {2008}),\ \Eprint{http://arxiv.org/abs/0803.0670}{arXiv:0803.0670 [hep-th]}%
  \bibAnnoteFile{NoStop}{Dutta:2008jt}%
\bibitem{Romanczukiewicz:2010eg}%
  \BibitemOpen
  \bibfield{author}{%
  \bibinfo {author} {\bibfnamefont{T.}~\bibnamefont{Romanczukiewicz}}\ and\
  \bibinfo {author} {\bibfnamefont{Y.}~\bibnamefont{Shnir}},\ }%
  \bibfield{journal}{%
  \Doi{10.1103/PhysRevLett.105.081601}{\bibinfo {journal} {Phys.Rev.Lett.}}\ }%
  \textbf{\bibinfo {volume} {105}},\ \bibinfo {pages} {081601} (\bibinfo {year}
  {2010}),\ \Eprint{http://arxiv.org/abs/1002.4484}{arXiv:1002.4484 [hep-th]}%
  \bibAnnoteFile{NoStop}{Romanczukiewicz:2010eg}%
\bibitem{Demidov:2011eu}%
  \BibitemOpen
  \bibfield{author}{%
  \bibinfo {author} {\bibfnamefont{S.}~\bibnamefont{Demidov}}\ and\ \bibinfo
  {author} {\bibfnamefont{D.}~\bibnamefont{Levkov}},\ }%
  \bibfield{journal}{%
  \Doi{10.1007/JHEP06(2011)016}{\bibinfo {journal} {JHEP}}\ }%
  \textbf{\bibinfo {volume} {1106}},\ \bibinfo {pages} {016} (\bibinfo {year}
  {2011}),\ \Eprint{http://arxiv.org/abs/1103.2133}{arXiv:1103.2133 [hep-th]}%
  \bibAnnoteFile{NoStop}{Demidov:2011eu}%
\bibitem{Anninos:1991un}%
  \BibitemOpen
  \bibfield{author}{%
  \bibinfo {author} {\bibfnamefont{P.}~\bibnamefont{Anninos}}, \bibinfo
  {author} {\bibfnamefont{S.}~\bibnamefont{Oliveira}},\ and\ \bibinfo {author}
  {\bibfnamefont{R.}~\bibnamefont{Matzner}},\ }%
  \bibfield{journal}{%
  \Doi{10.1103/PhysRevD.44.1147}{\bibinfo {journal} {Phys.Rev.}}\ }%
  \textbf{\bibinfo {volume} {D44}},\ \bibinfo {pages} {1147} (\bibinfo {year}
  {1991})%
  \bibAnnoteFile{NoStop}{Anninos:1991un}%
\bibitem{Campbell:1983xu}%
  \BibitemOpen
  \bibfield{author}{%
  \bibinfo {author} {\bibfnamefont{D.~K.}\ \bibnamefont{Campbell}}, \bibinfo
  {author} {\bibfnamefont{J.~F.}\ \bibnamefont{Schonfeld}},\ and\ \bibinfo
  {author} {\bibfnamefont{C.~A.}\ \bibnamefont{Wingate}},\ }%
  \bibfield{journal}{%
  \bibinfo {journal} {Physica}\ }%
  \textbf{\bibinfo {volume} {9D}},\ \bibinfo {pages} {1} (\bibinfo {year}
  {1983})%
  \bibAnnoteFile{NoStop}{Campbell:1983xu}%
\bibitem{Zamolodchikov:1978xm}%
  \BibitemOpen
  \bibfield{author}{%
  \bibinfo {author} {\bibfnamefont{A.~B.}\ \bibnamefont{Zamolodchikov}}\ and\
  \bibinfo {author} {\bibfnamefont{A.~B.}\ \bibnamefont{Zamolodchikov}},\ }%
  \bibfield{journal}{%
  \Doi{10.1016/0003-4916(79)90391-9}{\bibinfo {journal} {Annals Phys.}}\ }%
  \textbf{\bibinfo {volume} {120}},\ \bibinfo {pages} {253} (\bibinfo {year}
  {1979})%
  \bibAnnoteFile{NoStop}{Zamolodchikov:1978xm}%
\bibitem{Vachaspati:2006zz}%
  \BibitemOpen
  \bibfield{author}{%
  \bibinfo {author} {\bibfnamefont{T.}~\bibnamefont{Vachaspati}},\ }%
  \emph{\bibinfo {title} {{Kinks and domain walls: An introduction to classical
  and quantum solitons}}}\ (\bibinfo {year} {2006})%
  \bibAnnoteFile{NoStop}{Vachaspati:2006zz}%
\bibitem{Dashen:1975hd}%
  \BibitemOpen
  \bibfield{author}{%
  \bibinfo {author} {\bibfnamefont{R.~F.}\ \bibnamefont{Dashen}}, \bibinfo
  {author} {\bibfnamefont{B.}~\bibnamefont{Hasslacher}},\ and\ \bibinfo
  {author} {\bibfnamefont{A.}~\bibnamefont{Neveu}},\ }%
  \bibfield{journal}{%
  \Doi{10.1103/PhysRevD.11.3424}{\bibinfo {journal} {Phys.Rev.}}\ }%
  \textbf{\bibinfo {volume} {D11}},\ \bibinfo {pages} {3424} (\bibinfo {year}
  {1975})%
  \bibAnnoteFile{NoStop}{Dashen:1975hd}%
\bibitem{Teukolsky:1999rm}%
  \BibitemOpen
  \bibfield{author}{%
  \bibinfo {author} {\bibfnamefont{S.~A.}\ \bibnamefont{Teukolsky}},\ }%
  \bibfield{journal}{%
  \Doi{10.1103/PhysRevD.61.087501}{\bibinfo {journal} {Phys.Rev.}}\ }%
  \textbf{\bibinfo {volume} {D61}},\ \bibinfo {pages} {087501} (\bibinfo {year}
  {2000}),\ \Eprint{http://arxiv.org/abs/gr-qc/9909026}{arXiv:gr-qc/9909026
  [gr-qc]}%
  \bibAnnoteFile{NoStop}{Teukolsky:1999rm}%
\bibitem{Feroz:2008xx}%
  \BibitemOpen
  \bibfield{author}{%
  \bibinfo {author} {\bibfnamefont{F.}~\bibnamefont{Feroz}}, \bibinfo {author}
  {\bibfnamefont{M.}~\bibnamefont{Hobson}},\ and\ \bibinfo {author}
  {\bibfnamefont{M.}~\bibnamefont{Bridges}},\ }%
  \bibfield{journal}{%
  \Doi{DOI: 10.1111/j.1365-2966.2009.14548.x}{\bibinfo {journal}
  {Mon.Not.Roy.Astron.Soc.}}\ }%
  \textbf{\bibinfo {volume} {398}},\ \bibinfo {pages} {1601} (\bibinfo {year}
  {2009}),\ \Eprint{http://arxiv.org/abs/0809.3437}{arXiv:0809.3437
  [astro-ph]}%
  \bibAnnoteFile{NoStop}{Feroz:2008xx}%
\bibitem{Feroz:2007kg}%
  \BibitemOpen
  \bibfield{author}{%
  \bibinfo {author} {\bibfnamefont{F.}~\bibnamefont{Feroz}}\ and\ \bibinfo
  {author} {\bibfnamefont{M.}~\bibnamefont{Hobson}},\ }%
  \bibfield{journal}{%
  \Doi{10.1111/j.1365-2966.2007.12353.x}{\bibinfo {journal}
  {Mon.Not.Roy.Astron.Soc.}}\ }%
  \textbf{\bibinfo {volume} {384}},\ \bibinfo {pages} {449} (\bibinfo {year}
  {2008}),\ \Eprint{http://arxiv.org/abs/0704.3704}{arXiv:0704.3704
  [astro-ph]}%
  \bibAnnoteFile{NoStop}{Feroz:2007kg}%
\bibitem{PhysRevE.80.036603}%
  \BibitemOpen
  \bibfield{author}{%
  \bibinfo {author} {\bibfnamefont{M.}~\bibnamefont{Nishida}}, \bibinfo
  {author} {\bibfnamefont{Y.}~\bibnamefont{Furukawa}}, \bibinfo {author}
  {\bibfnamefont{T.}~\bibnamefont{Fujii}},\ and\ \bibinfo {author}
  {\bibfnamefont{N.}~\bibnamefont{Hatakenaka}},\ }%
  \bibfield{journal}{%
  \Doi{10.1103/PhysRevE.80.036603}{\bibinfo {journal} {Phys. Rev. E}}\ }%
  \textbf{\bibinfo {volume} {80}},\ \bibinfo {pages} {036603} (\bibinfo {month}
  {Sep}\ \bibinfo {year} {2009}),\
  \url{http://link.aps.org/doi/10.1103/PhysRevE.80.036603}%
  \bibAnnoteFile{NoStop}{PhysRevE.80.036603}%
\bibitem{Mandelstam:1975hb}%
  \BibitemOpen
  \bibfield{author}{%
  \bibinfo {author} {\bibfnamefont{S.}~\bibnamefont{Mandelstam}},\ }%
  \bibfield{journal}{%
  \Doi{10.1103/PhysRevD.11.3026}{\bibinfo {journal} {Phys.Rev.}}\ }%
  \textbf{\bibinfo {volume} {D11}},\ \bibinfo {pages} {3026} (\bibinfo {year}
  {1975})%
  \bibAnnoteFile{NoStop}{Mandelstam:1975hb}%
\end{thebibliography}%
\end{document}